\documentclass[a4paper,11pt]{article}
\usepackage[utf8]{inputenc}
\usepackage{mathtools}
\usepackage{natbib} 
\usepackage{graphicx}
\usepackage[font=footnotesize]{caption,subcaption} 
\usepackage{array} 
\usepackage{listings} 
\usepackage{lineno} 
\usepackage[nottoc,numbib]{tocbibind} 
\usepackage{hyperref} 
\usepackage[percent]{overpic} 
\usepackage{float} 
\usepackage{chngcntr} 
\usepackage{color}
\usepackage{authblk}
\usepackage{fullpage}
\usepackage{amsmath,amssymb,dsfont}
\usepackage{amsthm}
\usepackage{placeins}
\usepackage{tabularx}
\newcommand\upstrut{\rule{0pt}{12pt}}
\newcommand\downstrut{\rule[-6pt]{0pt}{6pt}}
\newcommand\mystrut{\upstrut\downstrut}

\newcolumntype{C}[1]{>{\centering\let\newline\\\arraybackslash\hspace{0pt}}m{#1}}

\renewcommand*{\thefootnote}{\fnsymbol{footnote}}
\newtheorem{theorem}{Theorem}
\newtheorem{proposition}[theorem]{Proposition}

\begin{document}

\title{Towards a taxonomy of learning dynamics in 2 $\times$ 2 games$^{\star}$}
\author[1]{Marco Pangallo}
\author[2]{James B. T. Sanders}
\author[2]{Tobias Galla}
\author[3,4,5]{J. Doyne Farmer}

\affil[1]{Institute of Economics and Department EMbeDS, 

Sant'Anna School of Advanced Studies, Pisa 56127, Italy}
\affil[2]{Theoretical Physics, School of Physics and Astronomy, University of Manchester, Manchester M13 9PL, UK}
\affil[3]{Institute for New Economic Thinking at the Oxford Martin School, University of Oxford, Oxford OX2 6ED, UK}
\affil[4]{Mathematical Institute, University of Oxford, Oxford OX1 3LP, UK}
\affil[5]{Santa Fe Institute, Santa Fe, NM 87501, US}

\date{\today}
\maketitle
\footnotetext[1]{Corresponding author: marcopangallo@gmail.com. For helpful comments and suggestions, we thank the Advisory Editor and two anonymous reviewers, as well as Vince Crawford,  Cars Hommes, Sam Howison, Peiran Jiao, Robin Nicole, Karl Schlag, Mihaela Van der Schaar, Alex Teytelboym, Peyton Young, and seminar participants at the EEA Annual Congress 2017, Nuffield College, INET YSI Plenary 2016, Herbert Simon Society International Workshop, Conference on Complex Systems 2016 and King’s College. Marco Pangallo performed the research presented in this paper when he was affiliated to the Institute for New Economic Thinking and Mathematical Institute  at the University of Oxford. He acknowledges financial support from INET and from the EPSRC award number 1657725.}

\abstract{Do boundedly rational players learn to choose equilibrium strategies as they play a game repeatedly? A large literature in behavioral game theory has proposed and experimentally tested various learning algorithms, but a comparative analysis of their equilibrium convergence properties is lacking. In this paper we analyze Experience-Weighted Attraction (EWA), which generalizes fictitious play, best-response dynamics, reinforcement learning and also replicator dynamics. Studying $2\times 2$ games for tractability, we recover some well-known results in the limiting cases in which EWA reduces to the learning rules that it generalizes, but also obtain new results for other parameterizations. For example, we show that in coordination games EWA may only converge to the Pareto-efficient equilibrium, never reaching the Pareto-inefficient one; that in Prisoner Dilemma games it may converge to fixed points of mutual cooperation; and that limit cycles or chaotic dynamics may be more likely with longer or shorter memory of previous play.}

\vspace{30pt}
\textbf{Key Words:} Behavioural Game Theory, EWA Learning, Convergence, Equilibrium, Chaos.

\textbf{JEL Class.:} C62, C73, D83.
 
\newpage
\renewcommand*{\thefootnote}{\arabic{footnote}}

\newpage
\section{Introduction}
\label{sec:Introduction}

In this paper we study boundedly rational players engaging in an infinitely repeated game. In this game, players update their stage-game strategies after every round using an adaptive learning rule.  We determine when players converge to a Nash Equilibrium (NE), when they converge to a stationary state that is not a NE, or when the learning dynamics never converge to any fixed point, asymptotically following either a limit cycle or a chaotic attractor.

More specifically, we analyze the learning dynamics of Experience-Weighted Attraction (EWA) \citep{camerer1999experience}. EWA is attractive for several reasons. From an experimental point of view, EWA has been shown to describe the behavior of real players relatively well in several classes of games, and is still widely used to model behavior in experiments. Our analysis, therefore, provides theoretical guidance on the learning dynamics that can be expected in experiments. From a theoretical point of view, EWA is attractive because it generalizes four well-known learning rules. Indeed, for some limiting values of its parameters, it reduces to best response dynamics, various forms of fictitious play \citep{fudenberg1998theory}, reinforcement learning and also a generalized two-population replicator dynamics with finite memory \citep{sato2003coupled}. Understanding the learning behavior under EWA makes it possible to generalize results about these four simpler learning algorithm by interpolating between the respective parameterizations. This yields new phenomena that may not be observed in the limiting cases.

We focus our analysis on 2-player games in which the same two players are repeatedly matched every time step to play the same stage game, which has two actions available per player. These are known as $2\times 2$ games.  We choose $2\times 2$ games because they encompass many of the strategic tensions that are typically studied by game theorists, and they are also simple enough to allow a comprehensive analytical characterization of the learning behavior under some parameterizations of EWA. While we are not able to provide a closed-form solution for all combinations of games and learning parameters, the parameterizations where we do provide a solution cover most previously studied cases and the transitions between them. We therefore go in the direction of providing a ``taxonomy'' of learning dynamics in $2\times 2$ games, for a family of learning rules and for any payoff matrix. 

In the limiting parameterizations at which EWA reduces to the learning rules that it generalizes, we recover well-known results. For example, our analysis shows that fictitious play always converges to one of the NE in $2\times 2$ games \citep{miyazawa1961convergence}. In particular, in Matching Pennies games, fictitious play converges to the mixed-strategy NE in the center of the strategy space, where players randomize between Heads and Tails with the same probability. On the contrary, in the limiting case at which EWA reduces to two-population replicator dynamics, it circles around the Matching Pennies equilibrium, which is also in line with the literature \citep{hofbauer1998evolutionary}. 

EWA parameters estimated from experimental data, however, rarely correspond to these limiting parameterizations, and are instead in the interior of the parameter space \citep{camerer1999experience}. This empirical fact makes it relevant to understand what happens for generic values of the parameters. Leaving the ``borders'' of the parameter space also yields several interesting new phenomena. For example, considering again Matching Pennies games, and the fictitious play and replicator dynamics learning rules, the role of memory in promoting convergence to equilibrium is not trivial. In fictitious play, longer memory makes convergence to equilibria more likely. Indeed, while the standard version of fictitious play, which has infinite memory, always converges to the mixed NE of Matching Pennies, a finite-memory version of fictitious play does not. Conversely, standard (two-population) replicator dynamics, which has infinite memory, does not converge to the mixed NE, while, we show, a finite memory generalization does. 

How is it possible that longer memory promotes equilibrium convergence in fictitious play, while it has the opposite effect in replicator dynamics? Our analysis of EWA learning makes sense of this difference, and identifies a precise boundary in the parameter space at which the effect of memory on stability flips sign. We show that it depends on the rate of growth of two key components of EWA, experience and attraction. When these two quantities grow at the same rate, as in fictitious play, players take a weighted average of previously experienced payoffs and new payoffs, and longer memory means that new payoffs are weighted less. So, longer memory intuitively promotes stability. Conversely, when experience does not grow or grows slower than attractions, longer memory does not imply that new payoffs are weighted less. In this case, it is shorter memory that promotes convergence, because quickly forgetting past payoffs makes it more likely that the players just randomize between their actions, without any player being strongly attracted towards Heads or Tails.

Another concrete example of the usefulness of going beyond the limiting cases of EWA is in understanding convergence to Pareto-inefficient equilibria in $2\times 2$ coordination games. Such games have two pure NE that can be Pareto-ranked. With simple learning rules such as fictitious play or replicator dynamics, the Pareto-inefficient NE is always locally stable. This means that if the players start sufficiently close to that equilibrium, they remain there forever. Our analysis shows that for certain values of the EWA parameters, and/or for very strong inefficiencies (i.e., the Pareto-optimal NE is clearly superior to the other NE), the Pareto-inefficient NE may cease to be locally stable. In other words, players would never remain stuck there, and always converge to the Pareto-optimal NE.\footnote{This outcome is similar to what could be expected based on stochastic stability \citep{young1993evolution}, but is obtained in a completely different framework.}

A final example concerns Prisoner Dilemma games. (In these games, our restriction to stage-game strategies may be less realistic than for the other games we study in this paper. Indeed, history-dependent strategies such as Tit-For-Tat have repeatedly been shown to be experimentally relevant.\footnote{Note that EWA could potentially model history-dependent strategies. For instance, \cite{galla2011cycles} considers a Prisoner Dilemma and three history-dependent strategies, always cooperate (AllC), always defect (AllD) and Tit-For-Tat (TFT). The stage game and the payoffs that these history-dependent strategies yield against each other define a game on which EWA can be run. We leave the study of such cases to future work. Moreover, we note that stage-game strategies may be more realistic in Prisoner Dilemmas if EWA is interpreted in a population dynamics sense: Every time step some player from a population plays a one-shot game against some randomly chosen player from another population. In this case, history-dependent strategies such as TFT are difficult as players do not know who they are going to play against.})   Under best response dynamics, fictitious play, and replicator dynamics, action profiles in which both players cooperate are never locally stable. This is because, under these three rules, players always consider forgone payoffs. If they start cooperating, when considering forgone payoffs they realize that, by unilaterally switching to defect, they may obtain higher payoffs. Under reinforcement learning, however, cooperation fixed points can be locally stable. This was shown by \cite{macy2002learning} under the name of stochastic collusion: Because in reinforcement learning players do not consider forgone payoffs, they do not realize that switching to defect yields better payoff, and so cooperation can be a stable outcome. Usefully, one of the EWA parameters interpolates between the extremes at which the players fully consider or ignore forgone payoffs. This makes it possible to precisely determine the point at which mutual cooperation ceases to be a stable outcome, depending on this parameter and on the payoffs.

From a practical point of view, our challenge is to characterize a 13-dimensional parameter space, composed of the eight payoffs that fully determine a $2\times 2$ game, the four EWA parameters, and the choice of the learning rule, which can be deterministic or stochastic (see below).\footnote{Strictly speaking, the stochasticity of the learning rule is not a parameter, but is still a dimension of our scenario analysis.} Our plan for the exploration of the parameter space is modular: We first consider a baseline scenario with a minimal number of free parameters, and then we study various extensions that involve varying the parameters that are fixed in the baseline scenario. Due to the strong non-linearity of EWA, we cannot provide a general closed-form solution for each parameter combination. However, we provide an example in which learning behavior in a part of the parameter space that we do not explicitly explore can be qualitatively understood on the basis of the scenarios that we study.

We start introducing the notation and defining the relevant classes of $2\times 2$ games in Section \ref{sec:classes2x2}. We then define the EWA learning rule in Section \ref{sec:EWA}. After that, in Section \ref{sec:overview} we give a qualitative overview of the main results, placing them in the context of the literature. This overview is more detailed than the one given in the introduction, and is meant to provide a deeper understanding of the results once the relevant notation has been introduced, without the need to dive into the technicalities of the mathematical analysis. We then discuss some simplifications that help the analysis and lay out a plan for the exploration of the parameter space in Section \ref{sec:preliminary}. Next, we analyze a baseline scenario in Section \ref{sec:baseline}, and we consider the dimensions of the parameter space that are not included in the baseline scenario in Section \ref{sec:stochbelief}. Section \ref{sec:conclusion} concludes. Most mathematical proofs are in the appendices, and additional results can be found in the Supplementary Appendix.

\section{Classes of 2 $\times$ 2 games}
\label{sec:classes2x2}

Despite being very simple, $2\times 2$ games encompass many of the strategic tensions that are studied by game theorists. In the following, we classify $2\times 2$ games into some classes that correspond to some of these strategic tensions, and that help understand the outcome of learning dynamics. 

We consider two-player, two-action games. We index the two players by $\mu \in \{\text{Row}=R,\text{Column}=C\}$ and write $s_i^\mu$ for the two actions of player $\mu$, with $i=1,2$. As usual we write $-\mu$ for the opponent of player $\mu$. When the two players choose actions $s_i^\mu$ and $s_j^{-\mu}$ player $\mu$ receives payoff $\Pi^\mu(s_i^\mu,s_j^{-\mu})$ and her opponent receives payoff $\Pi^{-\mu}(s_i^\mu,s_j^{-\mu})$. This can be encoded in a $2\times2$ bi-matrix of payoffs $\Pi$,
\begin{equation} 
\begin{tabular}{l|C{0.7cm}|C{0.7cm}|}
\cline{2-3}
                       &\upstrut $s_1^C$ \downstrut & \upstrut $s_2^C$ \downstrut \\ \hline
 \multicolumn{1}{|l|}{$s_1^R$} &\mystrut $a, e$ & $b, g$  \\ \hline
\multicolumn{1}{|l|}{$s_2^R$} & \mystrut $c, f$ & $d, h$  \\ \hline
\end{tabular}
\label{eq:payoff}
\end{equation}
where the element in position $(s_i^R,s_j^C)$ indicates the payoffs $\Pi^R(s_i^R,s_j^C),\Pi^C(s_i^R,s_j^C)$. For example, if the two players play actions $s_1^R$ and $s_2^C$, the payoffs are $b$ to player Row, and $g$ to player Column.

In the course of learning the two players can play mixed strategies, i.e. player $R$ plays action $s_1^R$ with probability $x$, and action $s_2^R$ with probability $1-x$. Similarly, player Column plays $s_1^C$ with probability $y$ and $s_2^C$ with probability $1-y$. The (time-dependent) strategy of player $R$ is encoded in the variable $x(t)$, and that of player Column by $y(t)$. Each of these variables is constrained to the interval between zero and one.

Based on the properties of the game one wants to look at, it is possible to construct several classifications of 2$\times$2 games. Perhaps the most famous classification was proposed by \cite{rapoport1966taxonomy}, who constructed all distinct games that can be obtained by ordering the payoffs of the two players in all possible ways. Our analysis below shows that, for many choices of the parameters, we do not need such a fine-grained classification of payoff matrices to build intuition into the outcome of EWA learning dynamics. It is instead enough to consider the pairwise ordering of the payoffs to one player when the action of the opponent is kept fixed. These comparisons are for example between $a$ and $c$ for player $R$, when the action of Column is fixed to $s_1^C$, and $b$ versus $d$ when Column's action is fixed to $s_2^C$. In principle there are $2^4=16$ such pairwise orderings. For our purposes we can group these orderings into 4 classes, as illustrated in Table~\ref{tab:gametypes}. These classes are also distinguished by the number, type and position of Nash equilibria.\footnote{Our classification is relatively standard, for example it is very close to the one in Chapter~10 of \cite{hofbauer1998evolutionary}.}
\\

\begin{table}[!t]
	\centering\small
		\begin{tabular}{ |>{\centering\arraybackslash} m{2.5cm}| >{\centering\arraybackslash} m{2.2cm}
		|>{\centering\arraybackslash} m{4.0cm}|>{\centering\arraybackslash} m{5.5cm}|}
		\hline
		Class of game & Payoffs &  Nash Equilibria & Example \\
   \hline
   		 Coordination & $a>c$, $b<d$,\newline $e>g$, $f<h$.   & Two pure strategy $(s_1^R,s_1^C)$, $(s_2^R,s_2^C)$ and one mixed strategy NE. & \vspace{3pt}  \includegraphics[width=0.32\textwidth]{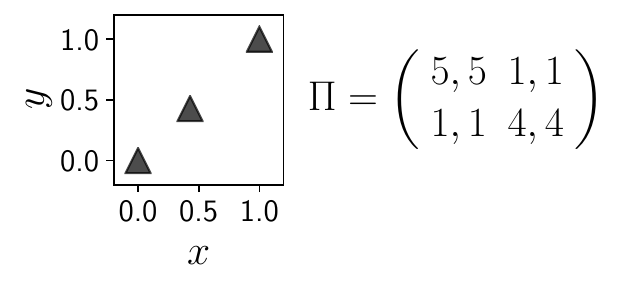} \\
			\hline
	 		 Anticoordination & $a<c$, $b>d$, \newline $e<g$, $f>h$.   & Two pure strategy $(s_1^R,s_2^C)$, $(s_2^R,s_1^C)$ and one mixed strategy NE. & \vspace{3pt}  \includegraphics[width=0.32\textwidth]{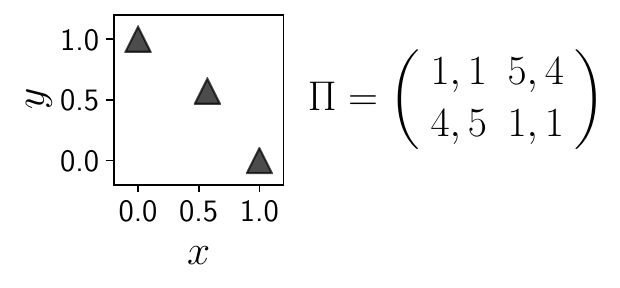} \\
			\hline
			 Cyclic &  $a>c$, $e<g$, \newline $b<d$, $f>h$; \newline $a<c$, $e>g$, \newline $b>d$, $f<h$.   & Unique mixed strategy NE. & \vspace{3pt}  \includegraphics[width=0.35\textwidth]{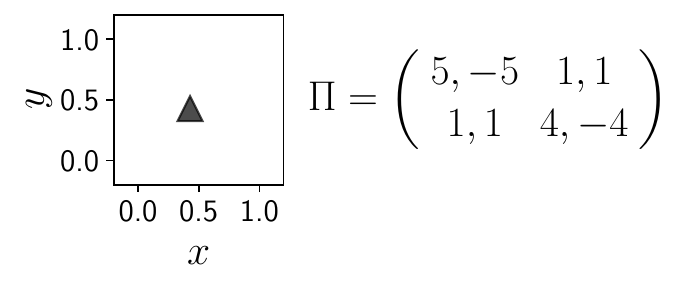}\\
			\hline		
			 Dominance-solvable & $a>c$, $e>g$, \newline $b>d$, $f>h$, \newline and all other \newline 11 orderings.  
	  & Unique pure strategy NE. & \vspace{3pt}  \includegraphics[width=0.35\textwidth]{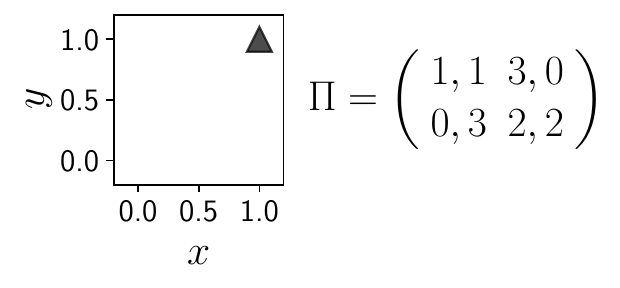} \\
			\hline			
		\end{tabular}
	\caption{Relevant classes of two-person, two-action games. The classes of games are defined from the pairwise ordering of the payoffs or, equivalently, from the number, type and position of Nash equilibria. For each class of games, we also provide an example payoff matrix $\Pi$ and the positions of the NE in the space defined by the probabilities $x$ and $y$ to play actions $s_1^R$ and $s_1^C$ respectively.}
	\label{tab:gametypes}
\end{table}

\noindent \textbf{Coordination and anticoordination games.} These correspond to the orderings $a>c$, $b<d$, $e>g$, $f<h$ (coordination games), and $a<c$, $b>d$, $e<g$, $f>h$ (anticoordination games).\footnote{\label{ftn_symm} Anticoordination and coordination games can be seen as equivalent, as each type of game can be obtained from the other by relabeling one action of one player (e.g. rename $s_1^R$  into $s_2^R$ and vice versa). However, part of our analysis will be based on symmetric games, and once the constraint of symmetry is enforced some properties of anticoordination and coordination games become different. Therefore, for clarity of exposition we keep coordination games distinct from anticoordination games.}  Coordination games have two pure strategy NE, one at  $(s_1^R,s_1^C)$ and the other at $(s_2^R,s_2^C)$. In addition there is one mixed strategy NE. Two well-known examples of coordination games are Stag-Hunt and Battle of the Sexes \citep{osborne1994course}. Anticoordination games also have two pure strategy and one mixed strategy NE, but at the pure strategy NE the players choose strategies with different labels, i.e. $(s_1^R,s_2^C)$ and $(s_2^R,s_1^C)$. A well-known example of an anticoordination game is Chicken.  
\\

\noindent \textbf{Cyclic games.} These correspond to the orderings $a>c$, $e<g$, $b<d$, $f>h$ or $a<c$, $e>g$, $b>d$, $f<h$. Games of this type are characterized by a cycle of best replies. For example, if one considers the first set of orderings, the best response to $s_1^R$ is for Column to play $s^C_2$. In response Row would choose $s_2^R$, Column would then play $s_1^C$, and the process would never converge to a fixed point. Cyclic games have a unique mixed strategy NE and no pure strategy NE. The prototypical cyclic game is Matching Pennies, which is a zero-sum game, but cyclic games in general need not be zero- or constant-sum. 
\\

\noindent \textbf{Dominance-solvable games.}  These comprise all $12$ remaining orderings. These games have a unique pure strategy NE, which can be obtained via elimination of dominated strategies. For instance, if $a>c$, $e>g$, $b>d$, $f>h$, the NE is $(s_1^R, s_1^C)$. The well-known Prisoners' Dilemma is a $2\times 2$ dominance-solvable game. (The dominance-solvable game shown in Table \ref{tab:gametypes} is a Prisoner Dilemma.)
\\

\section{Experience-Weighted Attraction learning}
\label{sec:EWA}

Experience-Weighted Attraction (EWA) has been introduced by \cite{camerer1999experience} to generalize two wide classes of learning rules, namely \textit{reinforcement} learning and \textit{belief} learning. Players using reinforcement learning are typically assumed to choose their actions based on the performance that these actions yielded in past play. Conversely, players using belief learning choose their actions by constructing a mental model of the actions that they think their opponent will play, and responding to this belief. \cite{camerer1999experience} showed that these two classes of learning rules are limiting cases of a more general learning rule, EWA. The connection lies in whether players consider \textit{forgone payoffs} in their update. If they do, for some parameters EWA reduces to belief learning. If they do not, it reduces to reinforcement learning. EWA interpolates between these two extremes and allows for more general learning specifications. 

We consider a game repeatedly played at discrete times $t=1,2,3,\ldots$. In EWA, players update two state variables at each time step. The first variable, $\mathcal N (t)$, is interpreted as \textit{experience}, as it grows monotonically as the game is played. The main intuition behind experience is that, the more the game is played, the less players may want to consider new payoffs obtained by playing certain actions, relative to their past experience with playing those same actions. The second variable, $Q_i^\mu(t)$, is the attraction that player $\mu$ has towards action $s_i^\mu$ (there is one attraction for each action). Attractions increase or decrease depending on whether realized or forgone payoffs are positive or negative. 

More formally, experience $\mathcal N (t)$ updates as follows:
\begin{equation}
\mathcal N (t) = \rho \mathcal N (t-1) + 1.
\label{eq:EWANt}
\end{equation}
In the above equation, each round of the game increases experience by one unit, although previous experience is discounted by a factor $\rho$. When $\rho=0$, experience never increases, while $\rho=1$ indicates that experience grows unbounded. For all other values $\rho \in (0,1)$, $\mathcal N (t)$ eventually converges to a steady state given by $\mathcal N^\star=1/(1-\rho)$.

Attractions are updated after every round of the game as follows:
\begin{equation}
	Q_i^\mu(t)=\frac{(1-\alpha) \mathcal N (t-1) Q_i^\mu(t-1)}{\mathcal N (t)}
 + \frac{\left[\delta + (1-\delta) I (s_i^\mu, s^{\mu}(t))\right] \Pi^\mu(s_i^\mu,s^{-\mu}(t))}{\mathcal N (t)}.
	\label{eq:EWA1}
\end{equation}
The first term discounts previous attractions. The memory-loss parameter $\alpha \in [0,1]$ determines how quickly previous attractions are discounted: when $\alpha=1$ the player immediately forgets all previous attractions, while $\alpha=0$ indicates no discounting. The second term in Eq. \eqref{eq:EWA1} is the gain or loss in attraction for action $s_i^\mu$. 

The term $\Pi^\mu(s_i^\mu, s^{-\mu}(t))$  is the payoff that player $\mu$ would have obtained from playing action $s_i^\mu$ against the action $s^{-\mu}(t)$ actually chosen by the other player at time $t$. We note that we have not specified whether $\mu$ has actually played $s_i^\mu$ or not. The parameter $\delta\in [0,1]$ controls how the attractions of $\mu$'s actions are updated, depending on whether player $\mu$ did or did not play a particular action.  The term $I (s_i^\mu, s^{\mu}(t))$ is the indicator function, and returns one if player $\mu$ played their action $s_i^\mu$ at time $t$, and zero otherwise. Therefore, if $\delta=1$, player $\mu$'s attractions for all of their actions are updated with equal weight, no matter what action $\mu$ played. That is, players take into account foregone payoffs. If, on the other hand, $\delta=0$, attractions are only updated for actions that were actually played. Intermediate values $0<\delta<1$ interpolate between these extremes.  

Irrespective of whether players consider forgone payoffs, the second term in Eq. (\ref{eq:EWA1}) is small when experience $\mathcal N (t)$ is large. This formalizes the intuition mentioned above, that players with a lot of experience may give little weight to newly experienced payoffs. In the updating of experience, Eq. \eqref{eq:EWANt}, we follow \cite{ho2007self} and redefine the parameter $\rho$ as $\rho=(1-\alpha)(1-\kappa)$. This redefinition is useful because the parameter $\kappa \in [0,1]$ makes it possible to more clearly interpolate between the various learning rules that EWA generalizes (see Section \ref{sec:rellit}). Because $\kappa$ determines $\rho$ once $\alpha$ is fixed, we refer to $\kappa$ as the discount rate of experience.\footnote{This is without loss of generality if $\kappa$ is unrestricted, because, except for $\alpha=1$, it is possible to obtain any value $\rho \in [0,1]$ by a suitable choice of $\kappa$. In the following, we will focus on $\kappa\in[0,1]$, but our analysis could be easily extended to general values of $\kappa$.} 

In EWA, the mixed strategies are determined from the attractions using a logit rule, see \cite{camerer1999experience}. For example, the probability for player Row to play pure strategy $s_1^R$ at time $t$ is given by
\begin{equation}
x(t)=\frac{e^{\beta Q_1^R(t)}}{e^{\beta Q_1^R(t)}+e^{\beta Q_2^R(t)}},
\label{eq:EWA0}
\end{equation}
and a similar expression holds for $y(t)$. The parameter $\beta\geq 0$ is often referred to as the intensity of choice in discrete choice models; it quantifies how much the players take into account the attractions for the different actions when they choose their actions. In the limit $\beta \rightarrow \infty$, for example, the players strictly choose the action with the largest attraction. For $\beta=0$, the attractions are irrelevant, and the players select among their actions randomly with equal probabilities.

\subsection{Special cases of EWA}
\label{sec:rellit}

\begin{figure}
\centering
\includegraphics[width=0.8\textwidth]{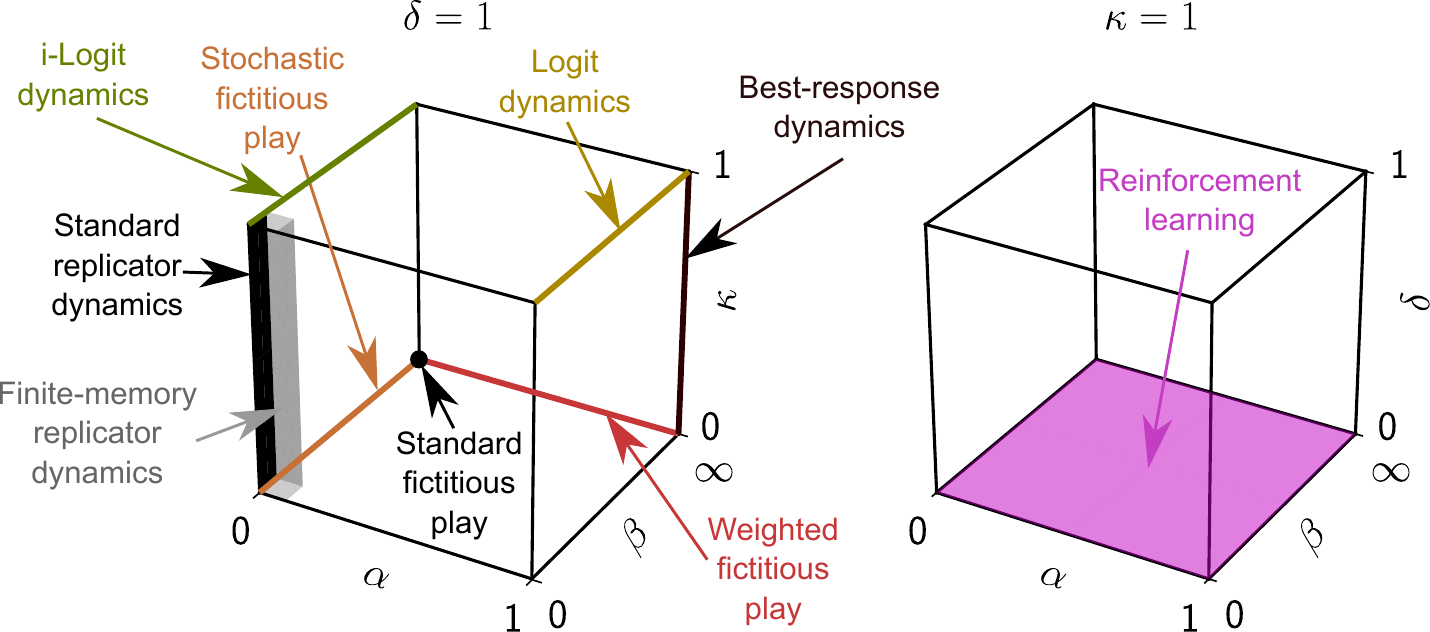}
\caption{Learning rules generalized by EWA. In this figure, on the left we show three EWA parameters: memory loss $\alpha$, payoff sensitivity $\beta$ and discount on experience $\kappa$. We fix the remaining parameter, the weight given to forgone payoffs $\delta$, to $\delta=1$. On the right, we fix $\kappa=1$ and show the remaining parameters $\alpha$, $\beta$, $\kappa$. See the main text for more details and a discussion on the learning rules. }
\label{fig:learningcube}
\end{figure}

Here, we present the parameter restrictions at which EWA reduces to the learning rules it generalizes (Figure \ref{fig:learningcube}).

When $\delta=0$, EWA reduces to reinforcement learning. In general, reinforcement learning corresponds to the idea that players update their attractions only considering the payoff they received, and so ignore forgone payoffs. Various specifications of reinforcement learning have been considered in the literature. For example, in \cite{erev1998predicting} attractions map linearly to probabilities, while \cite{mookherjee1994learning} consider the logit mapping in Eq. \eqref{eq:EWA0}. Depending on the value of $\kappa$, it is possible to have average reinforcement learning when $\kappa=0$, or cumulative reinforcement learning when $\kappa=1$. The difference between the two cases is that in average reinforcement players consider a weighted average of the payoff experienced in a given round and of past attractions, while in cumulative reinforcement they accumulate all payoffs without discounting past attractions.

The case $\alpha=1$, $\beta=+\infty$, $\delta=1$, for all values of $\kappa \in [0,1]$, is best response dynamics. Under best response dynamics, each player only considers her opponent's last move (complete memory loss of previous performance, $\alpha=1$), and plays her best response to that move with certainty ($\beta=+\infty$). Playing the best response generally requires fully taking into account the action that the player did not play in the previous round of the game ($\delta=1$). 

The case $\alpha=0$, $\beta=+\infty$, $\delta=1$ (and $\kappa = 0$) corresponds to \textit{fictitious play}. Differently from best response dynamics, players have infinite memory and best respond to the empirical distribution of actions of their opponent, which they take as an estimate of the opponent's mixed strategy. Fictitious play was proposed by \cite{brown1951iterative} and \cite{robinson1951iterative} as a method for finding the NE of a game and was later interpreted as a learning rule. Relaxing the assumption of infinite memory, the case with $\alpha\in (0,1)$ corresponds to \textit{weighted fictitious play}, as more recent history of play carries greater weight in estimating the opponent's mixed strategy. Conversely, if $\alpha=0$ but $\beta<+\infty$, the players do not choose with certainty the action with highest attraction and we have instead \textit{stochastic fictitious play}.\footnote{Finally, the combination of finite memory and finite intensity of choice, i.e. $\alpha \in (0,1)$, $\beta<+\infty$, $\delta=1$ (again with $\kappa=0$), results in weighted stochastic fictitious play.}

Both best-response dynamics and fictitious play are instances of belief learning, as in both cases players form beliefs about their opponent and respond to these beliefs.  This may not be apparent from Eq. \eqref{eq:EWA1}, which updates attractions in a way that more closely resembles reinforcement learning. Yet, \cite{camerer1999experience} show that the dynamics of \textit{expected} payoffs given beliefs is identical to the EWA dynamics as long as $\delta=1$ and $\kappa=0$. The first condition is intuitive: to compute expected payoffs, players need to consider both the actions that they played and the actions that they did not play. The second condition is more technical: it requires that attractions and experience are discounted at the same rate.\footnote{\cite{camerer1999experience} also discuss restrictions on the initial conditions of experience and attractions, $\mathcal N(0)$ and $Q_i^\mu(0)$. Initial conditions are also important to understand experimental play. In this paper we focus on the long-run dynamics of EWA for analytical tractability, so we do not stress the importance of initial conditions.}

Another learning dynamics that EWA generalizes is replicator dynamics. The limit $\beta \rightarrow 0$, with $\alpha=0$, $\delta=1$ and $\kappa\in(0,1]$, leads to \textit{two-population replicator dynamics} (see Supplementary Appendix \ref{sec:repldyn} for a derivation\footnote{Our derivation is different from \cite{borgers1997learning} and \cite{hopkins2002two} because these authors consider one of the versions of reinforcement learning proposed in \cite{erev1998predicting}, in which attractions map linearly to probabilities. We use instead a logit form. As a result, we get a different continuous time limit.}).  Assuming instead that $\alpha$ is positive but small, i.e. $\alpha \rightarrow 0$ (s.t. the ratio $\alpha/\beta$ is finite) we obtain a generalized \textit{two-population replicator dynamics with finite memory}, originally proposed by \cite{sato2003coupled}.

Finally, when $\alpha=1$, $\delta=1$ and $\kappa=1$, EWA is a discrete-time version of the so-called \textit{logit dynamics}; when $\alpha=0$, $\delta=1$ and $\kappa=1$, it reduces to the so called imitative or \textit{i-logit} dynamics. It should be noted, however, that both of these dynamics are generally studied in continuous time\citep{sandholm2010population}.

\section{Overview}
\label{sec:overview}

\begin{figure}[htbp]
\centering
\includegraphics[width=1\textwidth]{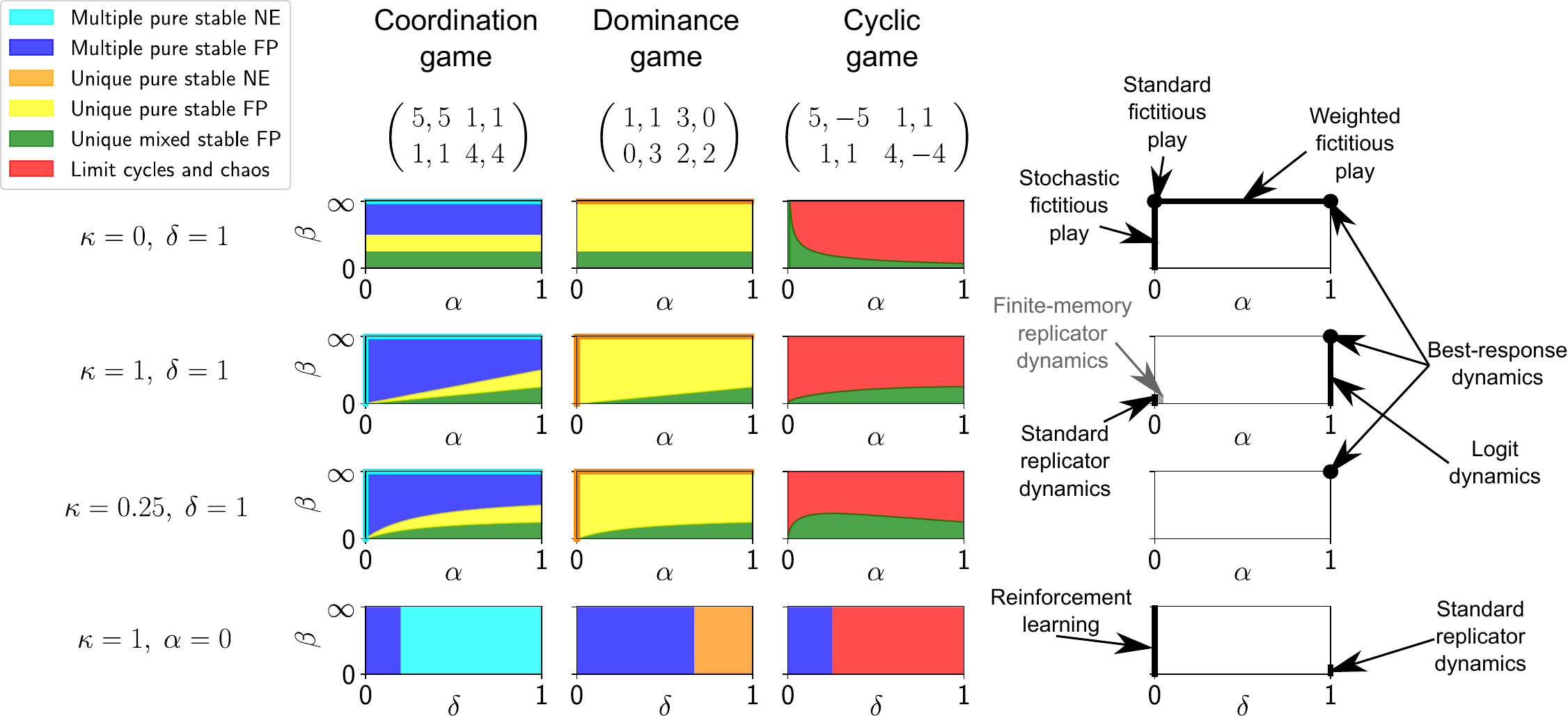}
\caption{Qualitative characterization of the outcome of learning under different parameters and games. We consider four cuts through the parameter space shown in Figure \ref{fig:learningcube}. In particular, we consider the planes defined by the restrictions $\kappa=0, \delta=1$; $\kappa=1, \delta=1$; $\kappa=0.25, \delta=1$; and $\kappa=1, \alpha=0$. For three games, we consider all possibilities for the asymptotic outcome of learning. (i) In cyan areas, learning converges to one of multiple pure NE; (ii) in blue zones, it converges to one of multiple fixed points that are located ``close'' to pure NE or at alternative pure strategy profiles; (iii) in orange areas, it converges to a unique pure strategy NE; (iv) in yellow zones, learning reaches a unique fixed point located close to a pure NE or at another pure strategy profile; (v) in green areas, it converges to a fixed point in the center of the strategy space; (vi) in red areas, it does not converge to any fixed point. On the right, we show the parameter restrictions at which EWA reduces to the algorithms that it generalizes.  }
\label{fig:summaryfigure}
\end{figure}

We proceed with an overview. Our goal is to give the reader a deeper understanding of the  results and of their place in the literature than in the introduction, without the need to dive into the technicalities of the mathematical analysis starting in Section \ref{sec:preliminary}. We discuss how leaving the borders of the parameter space in Figure \ref{fig:learningcube} gives new insights that would be missed when focusing on the learning algorithms that EWA generalizes.

We start from the case $\kappa=0$, $\delta=1$. This corresponds to the lower plane in the left panel of Figure \ref{fig:learningcube}, where EWA reduces to various forms of belief learning. Here, it is well-known that best-response dynamics always converges to pure strategy NE in all dominance-solvable games; it can converge to pure equilibria in coordination and anticoordination games, but depending on the initial condition it may also ``jump'' back and forth between the pure strategy profiles that are not equilibria; and it always cycles around the pure strategy profiles in cyclic games. Instead, fictitious play converges to (one of) the NE in all non-degenerate $2\times 2$ games \citep{miyazawa1961convergence,monderer1996a2}. This is no longer true in weighted fictitious play, as \cite{stahl1988instability} showed that this learning process does not converge in cyclic games, but it does converge to NE in all other $2\times 2$ games. Finally, in the case of stochastic fictitious play, learning converges to Quantal Response Equilibria \citep{mckelvey1995quantal,fudenberg1993learning,benaim1999mixed}, fixed points in the interior of the strategy space.\footnote{Weighted fictitious play has also been sparsely studied. \cite{cheung1997individual} experimentally test a population dynamics version of weighted stochastic fictitious play. They also show theoretically a transition between cycling and convergence to equilibrium for a certain value of the parameter $\beta$. \cite{benaim2009learning} study weighted stochastic fictitious play too, but take the limits $\alpha\rightarrow 0$, $\beta\rightarrow \infty$, and find convergence to a solution concept they propose.}

Our systematic characterization of the parameter space recovers all these results as special cases, precisely characterizing the position and stability of fixed points. For instance, in the case of stochastic fictitious play ($\alpha=0$) and coordination games, we show that the fixed point near the Pareto-inferior pure NE disappears as $\beta$ becomes small (transition from the blue to the yellow region), and eventually for very small $\beta$ the only stable fixed point is in the center of the parameter space.   Additionally, we show that for general values of $\alpha$ and $\beta$ (i.e., the interior of the plane), memory loss $\alpha$ does not determine the position and stability of fixed points in coordination and dominance games. However, it does determine stability of the unique mixed strategy fixed point in cyclic games. In particular, when $\alpha$ grows (i.e. players have shorter memory), for a given value of $\beta$, the fixed point is likely to become unstable. The fixed point also becomes unstable as $\beta$ grows. For these restrictions on $\kappa$ and $\delta$, shorter memory and more responsiveness to payoffs lead the players to cycle around the mixed strategy fixed point, without being able to spiral into it.

Another interesting case is the upper plane in Figure \ref{fig:learningcube} (left), corresponding to $\kappa=1$ and $\delta=1$. Here, EWA reduces to best-response dynamics, to the logit dynamics and to replicator dynamics. The logit dynamics reaches Quantal Response Equilibria in coordination and dominance games, and features a supercritical Hopf bifurcation in cyclic games \citep{sandholm2010population}. Standard (two-population) replicator dynamics is fully characterized in $2\times 2$ games \citep{hofbauer1998evolutionary} (Chapter 10). It converges to NE in all $2\times 2$ games, except in cyclic ones, where it circles around the unique mixed strategy NE. 

Again, our analysis reproduces these results. For instance, it is possible to see that the only fixed point at $\alpha=1$ loses stability for a finite value of $\beta$, suggesting that a Hopf bifurcation may be occurring. Our analysis also makes it possible to obtain new results. In coordination and dominance games, we see that the fixed point properties change as a linear function of $\alpha$ and $\beta$, i.e. it is the ratio $\alpha/\beta$ that matters. This is also true in the lower left rectangle of the $(\alpha,\beta)$ plane that represents replicator dynamics with finite memory. Moreover, this learning rule always converges to the unique mixed fixed point of cyclic games, in contrast with standard replicator dynamics that circles around the same fixed point (this result is not apparent from the figure, as one needs to take the limit $\alpha\rightarrow 0, \beta \rightarrow 0$, s.t. $\alpha/\beta$ is finite). Interestingly, in this case shorter memory makes it more likely that the fixed point is stable, in contrast with the case of weighted stochastic fictitious play ($\kappa=0$, $\delta=1$).

The effect of memory on stability becomes even more ambiguous in the $\kappa=0.25$, $\delta=1$ plane. Here, for a value of $\beta$ that is compatible with both the green and red regions (i.e. a horizontal line in the ($\alpha,\beta$) plane that cuts the boundary between the two regions twice), shorter or longer memory could make the unique mixed fixed point unstable. The inversion in slope of the function defining the boundary occurs precisely at $\alpha=\kappa=0.25$. In coordination and dominance games, the characteristics of fixed points are in between the $\kappa=0$ and $\kappa=1$ cases. 

Finally, we fix $\kappa=1$ and $\alpha=0$, and explore the $(\delta,\beta)$ plane. In this case, the parameter $\beta$ has no effect on fixed points, which are instead determined by $\delta$. In coordination games, for $\delta>1/5$, the EWA dynamics always converges to one of the two pure NE. However, when $\delta<1/5$, it can converge to fixed points corresponding to the two remaining pure strategy profiles. More interestingly, the same convergence to fixed points that are not NE occurs in the Prisoner Dilemma dominance-solvable game that we consider. For $\delta>2/3$, the only fixed point of the learning dynamics is the ($s_1^R,s_1^C$) action profile, which is also the unique NE of the game. This NE is Pareto inferior to ($s_2^R,s_2^C$), but players cannot coordinate on the Pareto-optimal action profile because they consider forgone payoffs for not deviating to $s_1^R$ or $s_1^C$. However, when $\delta<2/3$, they ignore forgone payoffs ``enough'' to make ($s_2^R,s_2^C$) a stable fixed point. A similar argument was given by \cite{macy1991learning} and \cite{macy2002learning}, who analyze the closely related Bush-Mosteller learning algorithm \citep{bush1955stochastic}, focusing on Prisoner Dilemma games. They introduce the concept of \textit{stochastic collusion}: Two players converge to a cooperation fixed point and keep cooperating because they do not realize that unilateral defection would be more rewarding. In a different context, our analysis reproduces this result. (As noted in the introduction, our result is most likely to be experimentally relevant if the learning dynamics is interpreted as representing one-shot interactions in a large population of learning agents.) A final point is that, in the cyclic game shown in Figure \ref{fig:summaryfigure}, learning converges to one of multiple pure strategy profiles when $\delta<0.25$. In other cyclic games, it may converge to one of a variety of fixed points located both on the edges and in the center of the strategy space (Section \ref{sec:reinforcement}).

\section{Preliminary steps}
\label{sec:preliminary}

In this section we prepare for our analysis of the outcomes of EWA learning. We first discuss a number of simplifications that help the analysis (Section \ref{sec:simpl}). We then introduce our plan for the exploration of the parameter space (Section \ref{sec:plan}).

\subsection{Simplifications}
\label{sec:simpl}

As a first simplification, we focus on the long-time outcome of learning, and assume that experience $\mathcal{N}(t)$ takes its fixed-point value $\mathcal N^\star = 1/(1-(1-\alpha) (1-\kappa))$. This assumption is valid at long times as long as $(1-\alpha) (1-\kappa) < 1$,\footnote{This restriction is always valid unless $\alpha=0$ and $\kappa=0$, as in standard and weighted fictitious play. However, it is possible to ex-post recover the convergence properties of these learning rules by taking the limit $\alpha \rightarrow 0$ in the stability analysis (see Section \ref{sec:belief}).} but in practice, for most values of the parameters, $\mathcal N^\star$ is reached after a few time steps.

Substituting the above fixed point into \eqref{eq:EWA1}, the update rule becomes
\begin{equation}
Q_i^\mu(t)=(1-\alpha) Q_i^\mu(t-1) + \left[1-(1-\alpha)(1-\kappa)\right] \left[\delta + (1-\delta) I (s_i^\mu, s^{\mu}(t))\right] \Pi^\mu(s_i^\mu, s^{-\mu}(t)).
\label{eq:EWA2}
\end{equation}

Our second simplification is to take a deterministic limit of the learning dynamics. Normally, learning dynamics are intrinsically stochastic. Indeed, when learning from playing a game repeatedly, during each round players can only observe one action of their opponent, and not her mixed strategy. The action that the opponent chooses is sampled stochastically from her mixed strategy vector, so the learning dynamics is inherently noisy. In this paper, by ``deterministic approximation'' we mean that the players play against each other an infinite number of times before updating their attractions, so that the empirical frequency of their actions corresponds to their mixed strategy. This sort of argument was already made by \cite{crawford1974learning} and justified by \cite{conlisk1993adaptation} in terms of fictitious ``two-rooms experiments'': The players only interact through a computer console and need to specify several actions before they know the actions of their opponent.\footnote{\cite{bloomfield1994learning} implemented this idea in an experimental setup. \cite{cheung1997individual} also consider a matching protocol and a population setting in which the players are matched with all players from the other population. This has a similar effect in eliciting mixed strategies.} This assumption is useful from a theoretical point of view and does not affect the results in most cases (Section \ref{sec:stoch}): the only difference when noise is allowed is a blurring of the dynamical properties.

We write $\overline{\Pi_i^R} (y(t))$ for the expected payoff to player Row from playing pure strategy $s_i^R$ at time $t$, given that player Column plays mixed strategy $y(t)$. For example, for $s_i^R=s_1^R$, the expected payoff is $\overline{\Pi_1^R}(y(t))=ay(t)+b(1-y(t))$. Similarly, we write $\overline{\Pi_j^C} (x(t))$ for the  expected payoff for player Column from playing action $s_j^C$, for a fixed mixed strategy $x(t)$ of player Row. The indicator function $I (s_i^\mu, s^{\mu}(t))$ can be replaced by the corresponding mixed strategy component, so for example $I (s_1^R, s^{R}(t)) \rightarrow x(t)$.

It is possible to combine Eqs. \eqref{eq:EWA0} and \eqref{eq:EWA2} and to formulate a closed map for $x(t), y(t)$. 

\begin{equation}
\begin{split}
x(t+1)=\frac{x(t)^{1-\alpha}e^{\beta \tilde{\kappa}[\delta + (1-\delta)x(t)] \overline{\Pi_1^R}(y(t))} }{x(t)^{1-\alpha}e^{\beta\tilde{\kappa}[\delta + (1-\delta)x(t)] \overline{\Pi_1^R}(y(t))}+(1-x(t))^{1-\alpha}e^{\beta \tilde{\kappa}[\delta + (1-\delta)(1-x(t))] \overline{\Pi_2^R}(y(t))}},\\
y(t+1)=\frac{y(t)^{1-\alpha}e^{\beta \tilde{\kappa}[\delta + (1-\delta)y(t)] \overline{\Pi_1^C}(x(t))} }{y(t)^{1-\alpha}e^{\beta \tilde{\kappa}[\delta + (1-\delta)y(t)] \overline{\Pi_1^C}(x(t))}+(1-y(t))^{1-\alpha}e^{\beta \tilde{\kappa}[\delta + (1-\delta)(1-y(t))] \overline{\Pi_2^C}(x(t))}},
\end{split}
\label{eq:EWA3}
\end{equation}
where $\tilde{\kappa}=1-(1-\alpha)(1-\kappa)$. 

We can obtain a continuous-time version of Eq. \eqref{eq:EWA3} by taking the limit $\alpha\rightarrow 0$ and $\beta\rightarrow 0$, such that the ratio $\alpha/\beta$ is finite. Further details can be found in Supplementary Appendix \ref{sec:repldyn}. When $\delta=1$ and $\alpha=0$, with $\beta \rightarrow 0$, EWA learning reduces to the standard form of the replicator dynamics. When $\alpha>0$ (although small), EWA reduces to a generalized form of the replicator dynamics with finite memory \citep{sato2003coupled,galla2013complex}.

Our third simplification is only valid when players fully consider forgone payoffs, i.e. $\delta=1$. In this case, it is possible to introduce a coordinate transformation that simplifies the dynamics, and helps to make the study of EWA analytically tractable. Specifically, we introduce the transformation\begin{equation}
\begin{split}
\tilde{x} = -\frac{1}{2} \ln{\left( \frac{1}{x}-1 \right)}, \\
\tilde{y} = -\frac{1}{2} \ln{\left( \frac{1}{y}-1 \right)},
\end{split}
\label{eq:coordinatetransformation}	
\end{equation}
only valid for $x,y$ within the interior of the strategy space, $x,y\in (0,1)$. Mathematically, this transformation of coordinates is a diffeomorphism; it leaves properties of the dynamical system such as Jacobian or Lyapunov exponents unchanged  \citep{ott2002chaos}. The original coordinates are restricted to $x(t)\in(0,1)$ and $y(t)\in(0,1)$, the transformed coordinates instead take values on the entire real axis. Pure strategies $(x,y) \in \{ (0,0),(0,1),(1,0),(1,1) \}$ in the original coordinates map to $(\tilde{x},\tilde{y}) \in \{ (\pm\infty,\pm\infty) \}$ in the transformed coordinates, with 0 mapping to $-\infty$ and 1 mapping to $+\infty$ (but for these values the transformation is not valid).  

In terms of the transformed coordinates (and assuming $\delta=1$), the map \eqref{eq:EWA3} reads
\begin{equation}
\begin{split}
\tilde{x}(t+1) = (1-\alpha) \tilde{x}(t) + \beta [1-(1-\alpha)(1-\kappa)] (A \tanh \tilde{y}(t) + B),\\
\tilde{y}(t+1) = (1-\alpha) \tilde{y}(t) + \beta [1-(1-\alpha)(1-\kappa)] (C \tanh \tilde{x}(t) + D),
\end{split}
\label{eq:EWA-newcoordinates}	
\end{equation}
where 
\begin{equation}
\begin{split}
A=\frac{1}{4} \left( a + d - b - c \right),\\
B=\frac{1}{4} \left( a + b - c - d \right),\\
C=\frac{1}{4} \left( e + h - f - g \right),\\
D=\frac{1}{4} \left( e + f - g - h \right).
\end{split}
\label{eq:newcoordinates-effparameters}	
\end{equation}
Eq. \eqref{eq:EWA-newcoordinates} underlines that, when $\delta=1$, the game only enters through the four payoff combinations $A$, $B$, $C$ and $D$. Broadly speaking, a positive value of the parameter $A$ indicates the preference of player Row for outcomes of the type $(s_1^R,s_1^C)$ or $(s_2^R,s_2^C)$ relative to outcomes $(s_1^R,s_2^C)$ and $(s_2^R,s_1^C)$. Similarly, positive values of $C$ indicate the  preference of player Column for the outcomes $(s_1^R,s_1^C)$ or $(s_2^R,s_2^C)$. These action combinations are the ones on the main diagonal of the payoff matrix. If instead $A$ is negative, player Row prefers off-diagonal combinations in the payoff matrix, and similarly player Column prefers off-diagonal combinations when $C$ is negative. The strength of these preferences for diagonal or off-diagonal combinations are determined by the modulus $|A|$ for player Row and by $|C|$ for player Column.  The parameter $B$ is a measure for the dominance of player Row's first action over her second, and similarly $D$ measures the dominance of player Column's first action over her second.  

The class of a $2\times 2$ game can also be established based on $A$, $B$, $C$ and $D$.
\begin{proposition}\label{prop0}
Consider a 2-player, 2-action game. The following statements hold:

\noindent (i)  The game is dominance solvable if $\left|B\right| > \left| A \right|$ or $\left|D\right| > \left| C \right|$;

\noindent (ii) If none of the conditions in (i) hold, and in addition $A>0, C>0$ the payoff matrix describes a coordination game; 

\noindent (iii) If none of the conditions in (i) hold, and in addition $A<0, C<0$ the payoff matrix describes an anticoordination game; 

\noindent (iv) If none of the conditions in (i) hold, and $A$ and $C$ have opposite signs (i.e. $AC<0$) the game is cyclic. 
\end{proposition}

To prove Proposition \ref{prop0}, it is sufficient to check that the restrictions on $A$, $B$, $C$ and $D$ translate into the inequalities in the second column of Table \ref{tab:gametypes}, and that viceversa the same inequalities imply the conditions on $A$, $B$, $C$ and $D$. We give a proof in Supplementary Appendix \ref{sec:proofprop0}.

\subsection{Plan for the exploration of the parameter space}
\label{sec:plan}

Our challenge is to characterize learning behavior in a 13-dimensional parameter space (the eight payoffs $a$, $b$, $c$, $d$, $e$, $f$, $g$, $h$; the four learning parameters $\alpha$, $\beta$, $\delta$ and $\kappa$; and the specification of the learning rule, which can be deterministic or stochastic). 

\begin{table}[htbp]
	\centering
		\begin{tabular}{|>{\centering\arraybackslash} C{2.5cm}|c|C{2cm}|c|c|c|c|c|c|}
			\hline
			 Case & Section & Payoffs & $\alpha$ & $\beta$ & $\delta$ & $\kappa$ & Rule & Analysis  \\
			\hline
			 Baseline & \ref{sec:baseline} & $A=\pm C$, $B=\pm D$ & --- & --- & $\delta=1$ & $\kappa=1$ & DET & AN-SIM \\
			\hline
			 Arbitrary payoffs & \ref{sec:asymmetric} & $A, B, C, D$ & --- & --- & $\delta=1$ & $\kappa=1$ & DET & AN \\
			\hline	
			 Belief learning & \ref{sec:belief} & $A, B, C, D$ & --- & --- & $\delta=1$ & --- & DET & AN \\
			\hline	
			 Reinforcement learning & \ref{sec:reinforcement} & $a,b,c,d$, $e,f,g,h$ & $\alpha = 1$ & --- & --- & $\kappa>0$ & DET & AN \\
			\hline	
			 Stochastic learning & \ref{sec:stoch} & $A, B, C, D$ & --- & --- & $\delta=1$ & $\kappa=1$ & STOCH & SIM \\
			\hline				
		\end{tabular}
		\caption{Plan for the exploration of the parameter space. Where the parameters are not set to any value (---), it means that in principle we fully analyze the dynamics for each value that the parameter can take. We start from a baseline scenario in which only four parameters are not fixed. We then follow a modular strategy, as in each case we explore the effect of varying one or more parameters at a time. For example, the case that we name ``arbitrary payoffs'' explores the effect of relaxing the constraints $A=\pm C$ and $B=\pm D$. In all cases with $\delta=1$ the payoffs can be reduced to the combinations $A$, $B$, $C$, $D$, so we indicate these parameters only; for the scenario that we name ``reinforcement learning'', this is not possible and so we indicate instead all payoffs $a,b,c,d,e,f,g,h$. The last two columns show whether the learning rule is stochastic (STOCH) or deterministic (DET), and whether our results are analytical (AN), obtained from simulations (SIM) or both (AN-SIM). }
	\label{tab:plan}
\end{table}

Due to non-linearity of EWA, we cannot obtain a closed-form characterization of the learning dynamics as a function of all parameter combinations. Therefore, we follow the modular strategy outlined in Table \ref{tab:plan}. In Section \ref{sec:baseline} we start with a baseline scenario in which dynamics is deterministic and only four parameters do not take a fixed value: these are the payoff combinations $A$ and $B$ ($C$ and $D$ are constrained to be either equal or of opposite sign than $A$ and $B$), the memory loss $\alpha$ and the intensity of choice $\beta$.  We consider this scenario as the baseline because it is the one with a minimal number of parameters, making it a clear benchmark against which to compare other parameterizations. Under the baseline scenario, we obtain most results analytically, either in closed-form or as the numerical solution of a fixed point equation. We also obtain some results by simulating the learning dynamics when no fixed points are stable.

We then consider various extensions, exploring the effect of changing one or more additional parameters while holding the others constant. For example, in Section \ref{sec:asymmetric} we relax the constraint that $A = \pm C$ and $B = \pm D$, and consider the effect of different combinations of payoffs to the two players. In Section \ref{sec:belief} we additionally relax $\kappa=1$ and fully explore the effect of changing $\kappa$ in the interval between 0 and 1 (the specific case $\kappa=0$ corresponds to belief learning). In Section \ref{sec:reinforcement} we let $\delta$ vary between 0 and 1, fixing $\alpha=1$ for analytical convenience (the specific case $\delta=0$ corresponds to reinforcement learning). In Section \ref{sec:stoch} we analyze stochastic learning, relaxing the deterministic approximation explained in Section \ref{sec:EWA}. While the results from most extensions are analytical, we study stochastic learning by simulations.

Why do we focus on these four extensions, while we could study many more, depending on the combinations of parameters that we vary and that we keep fixed? One reason is that we deem these four scenarios the most conceptually interesting. Another reason, we argue, is that it should be possible to qualitatively understand the learning behavior over the full parameter space as a superposition of the scenarios that we studied, which then can be considered as the most relevant. We give some argument for why this may be true in Section \ref{sec:otherpars}.

\section{Baseline scenario}
\label{sec:baseline}

We first analyze the asymptotic dynamics of EWA learning for the baseline scenario described in Table \ref{tab:plan}. In Section \ref{sec:fpanalysis} we analyze the existence and stability of fixed points, while in \ref{sec:simunstable} we simulate the learning dynamics in settings where all fixed points are unstable.  

\subsection{Fixed point analysis}
\label{sec:fpanalysis}

\subsubsection{Pure-strategy fixed points}
\label{sec:purestrat}

As can be seen in Eq. \eqref{eq:EWA3}, all pure strategy profiles are EWA fixed points. Intuitively, a pure strategy profile $i,j$ corresponds to infinite propensities $Q_i^R$ and $Q_j^C$, and finite changes in propensities (Eq. \ref{eq:EWA2}) have no effect. However, unless $\alpha=0$, all pure strategy fixed points are unstable. (If $\alpha=0$, only the Nash equilibria are stable pure strategy fixed points.) This is stated in the following proposition:

\begin{proposition}\label{prop1}
Consider a generic $2\times 2$ game and the EWA learning dynamics in Eq. \eqref{eq:EWA3}, with $\delta=1$ and $\kappa=1$. All profiles of pure strategies, $(x,y) \in \{ (0,0),(0,1),(1,0),(1,1) \}$ are fixed points of EWA. For positive memory loss, $\alpha > 0$, these fixed points are always unstable. When $\alpha=0$, the pure-strategy fixed points are stable if they are also NE, and unstable if they are not NE.
\end{proposition}
The proof of Proposition \ref{prop1} can be found in Appendix \ref{sec:proofprop1}.  

\subsubsection{Mixed-strategy fixed points in symmetric games}
\label{sec:symmetric}

EWA also has one or three mixed strategy fixed points, that is, fixed points in the interior of the strategy space. In the following, we characterize existence and stability of the mixed strategy fixed points. For convenience, we start from the case of symmetric games: this implies $A=C$ and $B=D$.\footnote{In a symmetric game the identity of the players does not matter, i.e. the payoff to player $\mu$ from playing action $s_i^\mu$ against action $s_j^{-\mu}$ does not depend on $\mu$. In formula, this means that $\Pi^R(s_i^R,s_j^C)=\Pi^C(s_j^R,s_i^C)$, so $A=C$ and $B=D$. We stress that in this paper symmetric games are just a special case to simplify the analysis, there is nothing else special about symmetry.} 

The location of the mixed strategy fixed points in the transformed coordinates, $( \tilde{x}^\star,\tilde{y}^\star)$ can be obtained from rearranging Eq. \eqref{eq:EWA-newcoordinates}. The fixed points are the solutions to $\tilde{x}^\star = \Psi^R(\tilde{x}^\star)$ and $\tilde{y}^\star= \Psi^C(\tilde{y}^\star)$, where
\begin{equation}
\begin{split}
\Psi^R(\tilde{x}^\star) = \frac{\beta}{\alpha} \left[ A \tanh \left( \frac{\beta}{\alpha} (C \tanh \tilde{x}^\star + D) \right) + B  \right], \\
\Psi^C(\tilde{y}^\star) = \frac{\beta}{\alpha} \left[ C \tanh \left( \frac{\beta}{\alpha} (A \tanh \tilde{y}^\star + B) \right) + D  \right].
\end{split}
\label{eq:fixedpoints-newcoordinates2}	
\end{equation}
It is already possible to note that the EWA parameters $\alpha$ and $\beta$ combine as the ratio $\alpha/\beta$. This justifies the linear shape of the transitions in the $(\alpha,\beta)$ plane of Figure \ref{fig:summaryfigure}, in coordination and dominance games (for $\kappa=1$, $\delta=1$).  Moreover, there is a scaling equivalence between increasing $\alpha/\beta$ or decreasing the payoff combinations $A$, $B$, $C$, $D$, as multiplying $\alpha/\beta$ by a constant and dividing the payoffs by the same constant leaves the fixed point equations unchanged.\\

\noindent {\em Fixed points and linear stability analysis}\\
Because in symmetric games $A=C$ and $B=D$, in turn $\Psi^R(\cdot) = \Psi^C(\cdot) = \Psi (\cdot)$. Depending on the class of the game and learning parameters, there can be either one or three mixed strategy fixed points.\footnote{While $\tilde{x}^\star=\Psi\left(\tilde{x}^\star\right)$ and $\tilde{y}^\star=\Psi\left(\tilde{y}^\star\right)$ implies that $\tilde{x}^\star$ and $\tilde{y}^\star$ take the same values, the \textit{pairs} $( \tilde{x}^\star,\tilde{y}^\star)$ are found by replacing the values in the original fixed point equations \eqref{eq:EWA-newcoordinates}. In particular, when there are three solutions to Eq. \eqref{eq:EWA-newcoordinates}, so that  $\tilde{x}^\star$ and $\tilde{y}^\star$ can take three values, the pairs $( \tilde{x}^\star,\tilde{y}^\star)$ need not be such that $\tilde{x}^\star=\tilde{y}^\star$.}

The Jacobian of the map in the transformed coordinates (obtained from Eq. \eqref{eq:EWA-newcoordinates}) is given by

\begin{equation}
J|_{\tilde{x}^\star,\tilde{y}^\star}= \begin{pmatrix}
  1 - \alpha & \frac{A \beta}{\cosh^2(\tilde{y}^\star)}  \\
  \frac{A \beta}{\cosh^2(\tilde{x}^\star)} & 1 - \alpha 
 \end{pmatrix},
\label{eq:symmetricstability}
\end{equation}
and its eigenvalues are
\begin{equation}
\lambda_\pm = 1- \alpha \pm \left| A \right| \beta \frac{1}{\cosh (\tilde{x}^\star) \cosh(\tilde{y}^\star)}.
\label{eq:symmetricstability-2}
\end{equation}
The fixed point is stable if $|\lambda_{\pm}|<1$. After some algebra this results in the stability condition
\begin{equation}
\frac{\alpha}{\beta}\cosh (\tilde{x}^\star) \cosh(\tilde{y}^\star) - \left| A \right| \geq 0.
\label{eq:symmetricstability-3}
\end{equation}

\noindent {\em Location and stability of fixed points}\\
We can now analyze the existence, location and stability of fixed points as we vary $A$ and $B$, while holding $\alpha/\beta=1$. (Again, up to this point only the combinations $(\beta/\alpha) A$ and $(\beta/\alpha) B$ matter, so changing the value of $\alpha/\beta$ is equivalent to rescaling the payoffs.)  It is in general not possible to obtain a closed-form solution for $\tilde{x}^\star$. Therefore, we first explore the parameter space by solving Eq. \eqref{eq:EWA-newcoordinates} numerically, and then provide some results for a number of limiting cases in which it is possible to obtain a closed-form solution.

\begin{figure}
\centering
\includegraphics[width=1\textwidth]{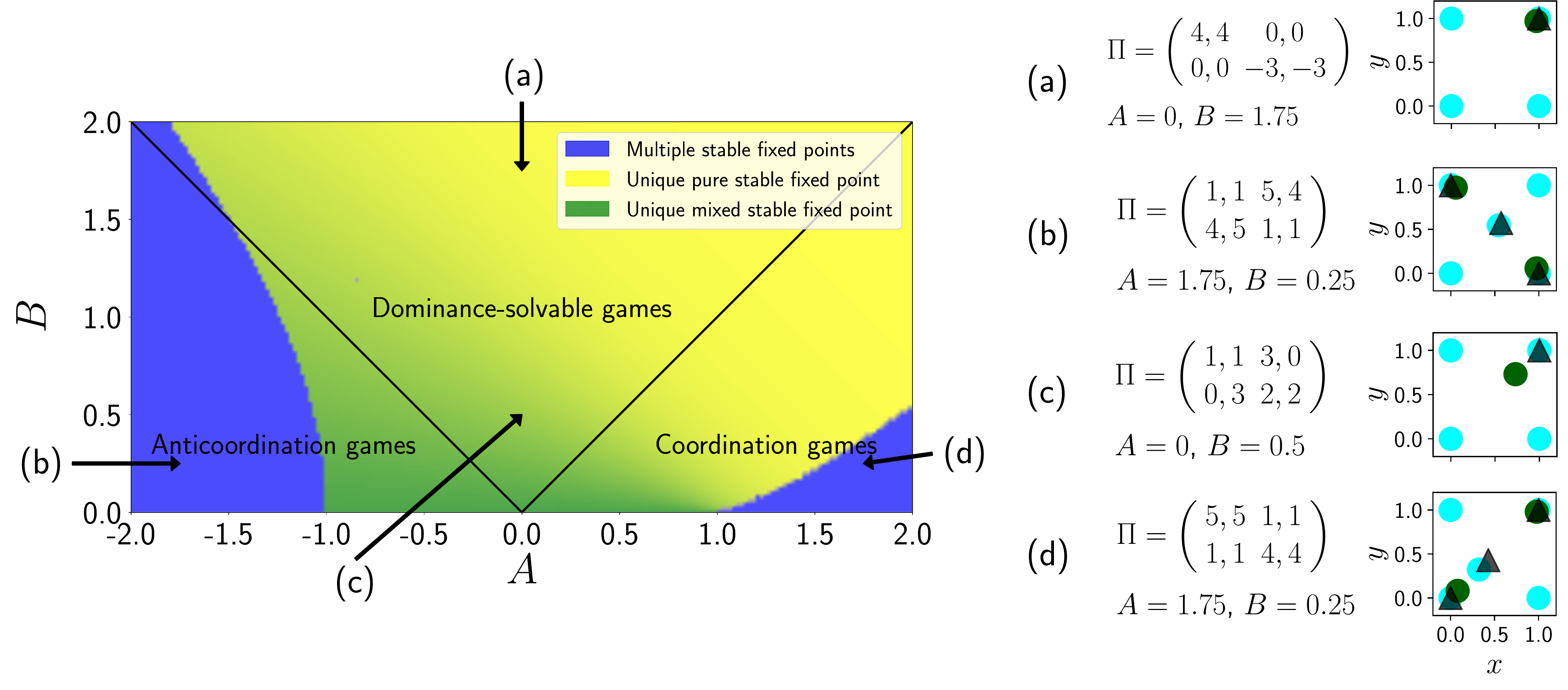}
\caption{Quantitative characterization of the parameter space in the special case $A=C$, $B=D$ (symmetric games), for fixed $\alpha/\beta=1$. The solid black lines in the figure separate the regions of anti-coordination games, dominance-solvable games and coordination games. The different colors are associated with different learning dynamics. In the blue region, there exist multiple stable (mixed-strategy) fixed points. In the green/yellow area there is only one stable fixed point. Through a linear interpolation the color gradient reflects the distance of the fixed point from the center of the strategy space: as the point in the $(A,B)$ plane becomes more yellow, the fixed point becomes closer to a pure strategy profile. The annotations from (a) to (d) on the borders refer to specific games shown on the right. For each game, we show its payoff matrix, the values of $A$ and $B$, and the position and stability of fixed points in the $(x,y)$ plane. Green circles are stable fixed points; cyan circles are unstable fixed points; grey triangles are NE.}
\label{fig:ACpositive}
\end{figure}

Figure \ref{fig:ACpositive} shows the properties of the fixed points as we vary $A$ and $B$, including a few typical examples.\\

\noindent{\em Unique fixed point near pure strategy NE:}\\
In case (a) of Figure \ref{fig:ACpositive} the payoff matrix describes a dominance-solvable game, in which actions $s_2^R$ and $s_2^C$ are strictly dominated by actions $s_1^R$ and $s_1^C$.  The fixed point is indicated by a green circle, and is located at $(x^\star,y^\star)=(0.95,0.95)$, very close to the unique pure strategy NE at $(1,1)$ (solid triangle). The fixed point is stable. As discussed in Section \ref{sec:purestrat}, all pure-strategy profiles are unstable fixed points (cyan circles). 
\\

\noindent{\em Multiple stable fixed points near pure-strategy NE:}\\
Cases (b) and (d) are examples of anticoordination and coordination games respectively. Each of the two games has three NE, as is indicated by the triangles. In each example one NE involves a mixed strategy, and the other two equilibria are pure-strategy NE. In both cases the values of $A$ and $B$ are such that there are three fixed points of EWA learning. For both examples there exist a ``central'' fixed point, located near the mixed-strategy NE and unstable under the learning dynamics (cyan circle near the centre of strategy space), and two stable ``lateral'' fixed points.  

The important difference between the two cases is that in (b) both pure-strategy NE are also Pareto equilibria, whereas in (d) $(s_1^R,s_1^C)$ is both a NE and a Pareto equilibrium. 
This generates the asymmetry between the $A>0$ and $A<0$ semiplanes.\footnote{Note that the discrepancy between coordination and anticoordination games here is an artifact of the symmetry assumption $A=C$ and $B=D$. Indeed, an anticoordination game with payoff matrix $ \left( {\begin{array}{cc}
   1,1 & 5,5 \\       4,4 & 1,1   \end{array} } \right)$ is asymmetric, but perfectly equivalent to case (d) in terms of Pareto-efficiency. See also footnote \ref{ftn_symm}.} When $A>0$ and $B$ gets larger, the payoff discrepancy between the Pareto-efficient NE and the Pareto-inefficient NE increases.  The stable lateral fixed point closest to the Pareto-inefficient NE collides with the unstable central fixed point, generating a \textit{fold bifurcation} in which both fixed points disappear. Effectively, positive memory loss and non-infinite payoff sensitivity prevent the learning dynamics from getting stuck in a ``local minimum'',
and help reaching the Pareto-efficient NE.
\\

\noindent{\em Unique fixed point away from pure-strategy NE:}\\
Case (c) corresponds to a dominance game like (a), but the payoffs are smaller than in the previous example. As the payoffs are smaller, there are less incentives to learn: the only stable fixed point of EWA learning is closer to the centre of strategy space than in case (a).   \\

\noindent {\em Analytical results:}\\
We proceed with some analytical results for a number of specific cases. We first set $B=0$. The boundaries between the blue and green areas in Fig. \ref{fig:ACpositive} is then found at $A=-1$ and $A=1$. Mathematically, these boundaries mark the point at which the lateral fixed points cease to exist (they are present in the blue areas, but not in the green area). Calculating the slope of $\Psi(u)$ at $u=0$, one shows that the lateral fixed points do not exist if  
\begin{equation}
\frac{\beta}{\alpha} \left|A\right| \leq 1,
\label{eq:symmetricexistence1}
\end{equation}
leading immediately to the boundaries $A=\pm 1$ in Figure \ref{fig:ACpositive} ($\alpha/\beta=1$ in the figure).

When $\frac{\beta}{\alpha} \left|A\right| \rightarrow +\infty$, $\Psi(\tilde{x}^\star)$ approaches a step function equal to $-\frac{\beta}{\alpha} \left|A\right|$ in the negative domain and to $\frac{\beta}{\alpha} \left|A\right|$ in the positive domain, so the intersections with the $\tilde{x}^\star$ line occur precisely at $\tilde{x}^\star=0$ and $\tilde{x}^\star=\pm\frac{\beta}{\alpha} \left|A\right|$.  Recalling the mapping from the transformed coordinates to the original coordinates, these intersections correspond to $x=0$, $x=1/2$ and $x=1$. By using the same argument for $y$, it is easy to see that the fixed points are the pure strategy NE of the coordination/anticoordination game and the mixed equilibrium in the center of the strategy space. In Figure \ref{fig:ACpositive}, cases (b) and (d) approximate this situation.

We now consider $B\neq 0$.  If $\frac{\beta}{\alpha} \left|B\right| \rightarrow +\infty$ and $B\gg A$, $\Psi(\tilde{x}^\star)$ is completely flat and equal to $\Psi(0)=\frac{\beta}{\alpha}B$. This is also the position of the unique fixed point $\tilde{x}^\star$. As $\tilde{x}^\star\rightarrow \pm \infty$ (depending on the sign of $B$), $x\rightarrow 0,1$ and the fixed point corresponds to the unique pure strategy NE.   Case (a) in Figure \ref{fig:ACpositive} approximates this situation.

Stability is addressed in the following proposition.

\begin{proposition}\label{prop2}
In symmetric $2\times 2$ games and with the learning parameters taking values as in the baseline scenario (Table \ref{tab:plan}) the following results hold:

(i) if $B=0$ and $\frac{\beta}{\alpha} \left|A\right| \leq 1$, the unique central fixed point is stable.

(ii) if $B=0$ and $\frac{\beta}{\alpha} \left|A\right| \rightarrow 1^+$ or $\frac{\beta}{\alpha} \left|A\right| \rightarrow +\infty$, the central fixed point becomes unstable and the lateral fixed points are stable. In particular, at $\frac{\beta}{\alpha} \left|A\right| = 1$ a supercritical pitchfork bifurcation occurs. 

(iii) if $\frac{\beta}{\alpha} \left|B\right| \rightarrow +\infty$ and $B \gg A$, the unique fixed point is stable. 
\end{proposition}

The proof is in Appendix \ref{sec:proofprop2}. In sum, in symmetric $2\times 2$ games at least one fixed point is always stable, at least in the limiting cases covered in the proposition (but the numerical analysis above suggests that the results in the proposition also hold for intermediate values).

\subsubsection{Mixed-strategy fixed points in asymmetric games}
\label{sec:antisymmetric}

We focus on a specific type of asymmetric games in which the asymmetry only stems from the sign of the payoffs. These games are defined by the condition $\Pi^R(s_i^R,s_j^C)=-\Pi^C(s_j^R,s_i^C)$, which implies $A=-C$, $B=-D$. Note that this condition does not generally define zero-sum games, which are rather defined by the equality $\Pi^R(s_i^R,s_j^C)=-\Pi^C(s_i^R,s_j^C)$.\footnote{These asymmetric games and zero-sum games only correspond if  $\Pi^R(s_i^R,s_j^C)=\Pi^C(s_i^R,s_j^C)=0$ for $i\neq j$.} Under this definition, if $B>A$ the game is dominance-solvable, but if $A>B$ we have a cyclic game. 
\\

\noindent{\em Fixed points and stability:}

As in the previous section, we first write down the conditions for the existence and stability of fixed points, and then study their properties as we vary the learning parameters and the payoffs. 

When $A=-C$, Eqs. \eqref{eq:fixedpoints-newcoordinates2} have at most one solution, as the functions on the right hand side monotonically decrease. Moreover, if $B\neq 0$ we generally have $\tilde{x}^\star \neq \tilde{y}^\star$. The eigenvalues of the Jacobian \eqref{eq:symmetricstability} are complex and of the form
\begin{equation}
\lambda_\pm = 1- \alpha \pm i  \frac{\beta\left| A \right|}{\cosh (\tilde{x}^\star) \cosh(\tilde{y}^\star)}.
\label{eq:antisymmetricstability-2}
\end{equation}
The stability condition is then:\footnote{Here we just find the condition under which $(\tilde{x}^\star,\tilde{y}^\star)$ --- the only potentially stable fixed point --- loses stability. It is possible to prove that the dynamical system undergoes a supercritical Hopf bifurcation (or Neimark-Sacker bifurcation) when the eigenvalues cross the unit circle. However, the proof involves calculating the so-called first Lyapunov coefficient, which requires a lot of algebra and does not provide any insight, so we do not provide a proof here. We instead use numerical simulations to show that the Hopf bifurcation is indeed supercritical.}
\begin{equation}
\frac{\beta}{\sqrt{2\alpha-\alpha^2}}\frac{\left| A \right|}{\cosh (\tilde{x}^\star) \cosh(\tilde{y}^\star)} \leq 1.
\label{eq:antisymmetricstability-3}
\end{equation}
This stability condition is different from the one of symmetric games, in Eq. \eqref{eq:symmetricstability-2}. Indeed, it is not only the ratio $\alpha/\beta$ that matters, but a more complicated function of these parameters. In general, increasing $\alpha$ or $\beta$ has the same effect on stability as with the ratio $\alpha/\beta$, but when taking the limit $\alpha,\beta\rightarrow 0$ (such that the ratio $\alpha/\beta$ is finite), the left hand side of the above equation goes to zero, and so the fixed point is always stable. This is consistent with replicator dynamics with finite memory always converging to a mixed strategy fixed point (see Supplementary Appendix \ref{sec:repldyn}), which could however be arbitrarily far from a Nash equilibrium.
\\

\noindent{\em Examples of typical behaviour:}

\begin{figure}[h]
\centering
\includegraphics[width=1\textwidth]{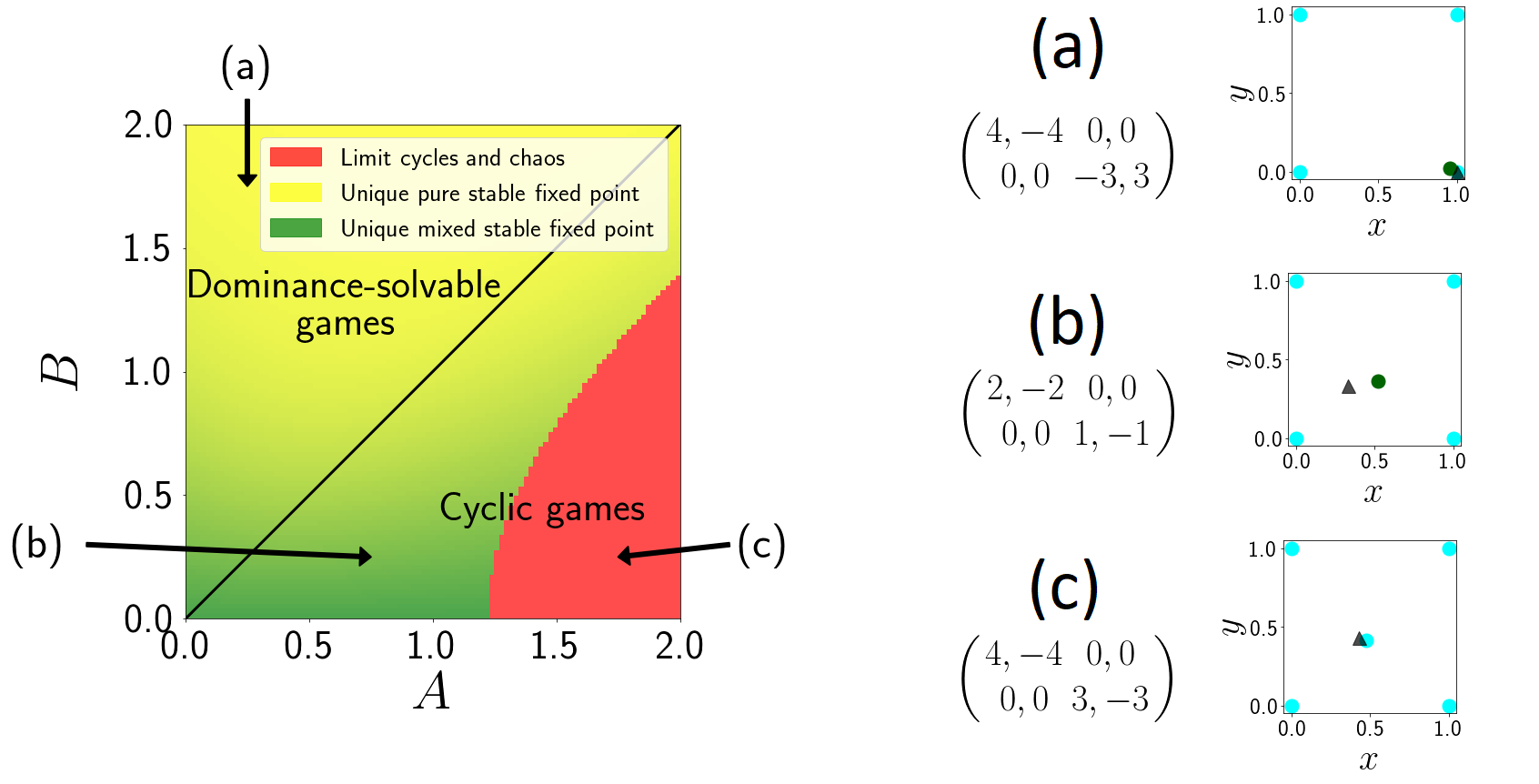}
\caption{Quantitative characterization of the parameter space of asymmetric games in which $A=-C$ and $B=-D$, for $\alpha=\beta=0.8$. This figure has the same interpretation as Figure \ref{fig:ACpositive}. In the red portion of the parameter space no fixed points are stable, and the learning dynamics follows limit cycles or chaos.}
\label{fig:ACnegative}
\end{figure}

In Figure \ref{fig:ACnegative} we illustrate the different possible outcomes for asymmetric games, as we did for symmetric games in Fig. \ref{fig:ACpositive}. Example (a) is a dominance-solvable game. The learning dynamics converges to a unique fixed point close to the pure strategy NE, analogously to case (a) in Figure \ref{fig:ACpositive}. In case (b) we have instead a cyclic game, with relatively low values of the payoffs. As in symmetric games, low values of the payoffs imply that the fixed point at the center of strategy space --- not necessarily corresponding to the NE --- is stable. Case (c) is similar to case (b), but the payoffs are larger. Higher incentives make the players overreact to their opponent's actions, and this makes all fixed points unstable. The learning dynamics gets trapped in limit cycles or, for some parameters, chaotic attractors, as we will show in Section \ref{sec:simunstable}.

\vspace{1em}
\noindent {\em Cyclic games -- Matching Pennies:}

We next focus on a specific example of cyclic games, Matching Pennies. This is a zero-sum game in which one player gains a coin, while the other player loses the coin \citep{osborne1994course}. The resulting payoff matrix implies $B=D=0$, $C=-A$. The learning dynamics have a unique fixed point at $(\tilde{x}^\star,\tilde{y}^\star)=(0,0)$. Replacing in Eq. \eqref{eq:antisymmetricstability-3} we find that the fixed point is stable if 
\begin{equation}
\frac{\beta}{\sqrt{2\alpha - \alpha^2}}   |A| \leq 1.
\label{eq:asymmetricstability-2}
\end{equation}
For the values of $\alpha$ and $\beta$ used in Figure \ref{fig:ACnegative}, the fixed point becomes unstable for $A^\star=1.224$. This corresponds the the boundary of the green and red areas for $B=0$, at the bottom of Fig. \ref{fig:ACnegative}.
\\

Summing up, in asymmetric games defined by the constraint $A=-C$ and $B=-D$, there exists one stable fixed point unless $A>B$, in which case the fixed point may lose stability.

\subsection{Simulations of unstable dynamics}
\label{sec:simunstable}

All analysis so far was about local stability of fixed points. We now simulate dynamics to assess global stability and to check which type of dynamics arise when all fixed points are unstable.

In symmetric games, dynamics always converge to one of the stable fixed points, except in one case. When $\beta$ is large, $\alpha$ is small and $|A|\gg |B|$ (coordination or anticoordination game), for some initial condition close to the action profiles that are not NE, it is possible to observe a stable limit cycle of period 2. In this cycle the players ``jump'' between the pure strategy profiles that are not NE of the coordination/anticoordination game. This is unsurprising, as these parameter restrictions make EWA closely resemble best response dynamics (see Section \ref{sec:rellit}). As this dynamics is behaviorally unrealistic and not robust to stochasticity --- it is enough that one player ``trembles'' and the dynamics converges to the NE --- we ignore it for the rest of the analysis. It is just an artifact of the deterministic approximation.

\begin{figure}[h]
\centering
\includegraphics[width=0.7\textwidth]{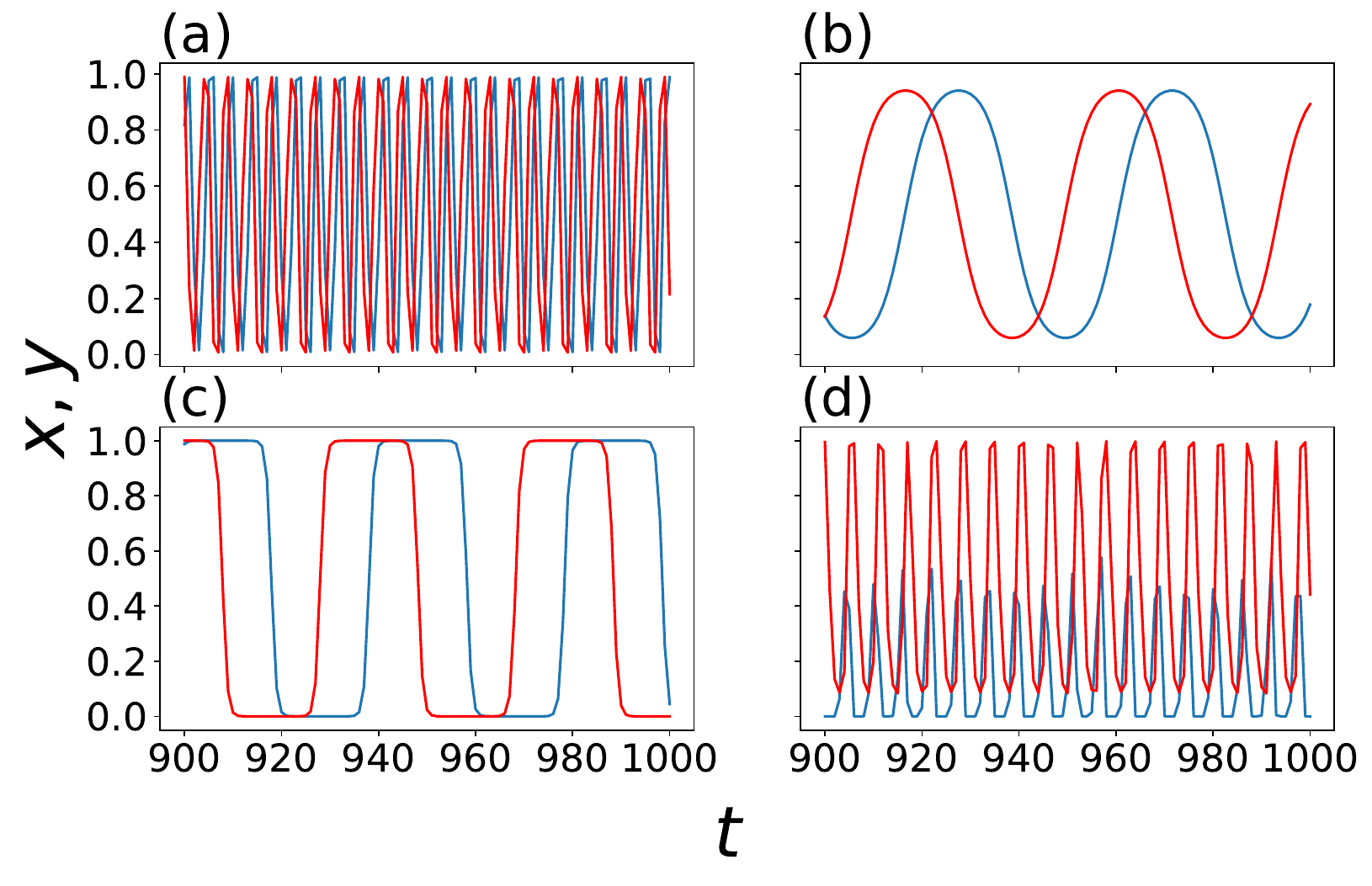}
\caption{Time series of the probabilities $x$ (in blue) and $y$ (in red), for four different combinations of learning parameters and payoffs (detailed in the text).  Cyclical and chaotic dynamics occur.}
\label{fig:TimeSeries_asymmetricgames_limitcycle}
\end{figure}

In asymmetric games in which all EWA fixed points are unstable, we instead observe more behaviorally realistic strategic oscillations. To illustrate the nature of the unstable solutions, Figure \ref{fig:TimeSeries_asymmetricgames_limitcycle} shows some examples of the learning dynamics for some values of $\alpha$, $\beta$, $A$ and $B$.  In panels (a) to (c) we have $A=-C=2$ and $B=D=0$, while in panel (d) we consider $A=-C=-3.4$ and $B=-D=-2.5$.

In panel (a), for $\alpha=0.7$ and $\beta=1$, the players frequently change their strategies, whereas in panel (b), for $\alpha=0.01$ and $\beta=0.1$, the dynamics is smoother.  Note that the ratio $\alpha/\beta$ is very similar in the two cases, but nonetheless the dynamics is quite different. This is not in contradiction with the rest of the paper: only the fixed point behavior of EWA is determined by the ratio $\alpha/\beta$. In panel (c), where $\alpha = 0.01$ and $\beta = 0.5$ the players spend a lot of time playing mostly one action and then quickly switch to the other action (because they have long memory and high payoff sensitivity). Finally, in panel (d), we choose $B \neq 0$: this seems to yield the most irregular dynamics. In Supplementary Appendix \ref{sec:chaoticdynamics}, we show that these dynamics are chaotic.\\

\section{Extensions}
\label{sec:stochbelief}

We now consider the extensions to the baseline scenario (see Table \ref{tab:plan}). In Section \ref{sec:asymmetric} we consider games in which payoffs are not constrained by $A=\pm C$ and $B=\pm D$, so that the magnitude of the payoffs can be different to the two players. In Section \ref{sec:belief} we consider values of the parameter $\kappa \in [0,1)$ (in the case $\kappa=0$ we recover belief learning). In Section \ref{sec:reinforcement} we consider $\delta\in[0,1)$, recovering reinforcement learning for the case $\delta=0$. In Section \ref{sec:stoch} we drop the simplification of deterministic learning and analyze the stochastic learning dynamics. 

These extensions do not cover all the 13-dimensional parameter space described in Section \ref{sec:plan}. As discussed elsewhere, it is beyond the reach of this paper to fully explore the parameter space; the previously considered regions cover a lot of interesting transitions between the learning algorithms that EWA generalizes. Yet, in Supplementary Appendix \ref{sec:otherpars}, we consider a few parameter and payoff combinations that have not been explicitly covered in the previous analysis. We show that for the specific games and payoffs considered, we are able to qualitatively understand the learning dynamics based on the baseline scenario and on the scenarios studied in this section. While we cannot claim that this is true in general, we consider this an encouraging sign.

\subsection{Arbitrary payoffs}
\label{sec:asymmetric}

From the point of view of learning, games in which $A\neq C$ and $B\neq D$ are widely similar to games in the same class for which the constraint $A=\pm C$ and $B=\pm D$ holds. For example, dominance-solvable games with arbitrary payoffs are widely similar to dominance-solvable games with constrained payoffs. In Supplementary Appendix \ref{sec:suppasymmgames}, we show a few examples in which payoffs to one player are larger than payoffs to the other player, leading the player with highest payoffs to play mixed strategies closer to the boundaries of the strategy space.

Our same analytical results in Proposition \ref{prop1} and Eq. \eqref{eq:fixedpoints-newcoordinates2} apply, and stability can be obtained replacing $|A|\rightarrow \sqrt{AC}$ in Eq. \eqref{eq:symmetricstability-3} when $AC>0$, and in Eq. \eqref{eq:antisymmetricstability-3} when $AC<0$.

\subsection{Belief learning}
\label{sec:belief}

Choosing $\kappa\neq 1$ in   Eqs. \eqref{eq:EWA3} and \eqref{eq:EWA-newcoordinates} is equivalent to rescaling the payoff sensitivity $\beta$ as follows
\begin{equation}
\tilde{\beta}=\beta \left[1-(1-\alpha)(1-\kappa) \right].
\label{eq:betatilde}
\end{equation}
As the quantity multiplying $\beta$ is smaller than one for $\kappa < 1$, the effective payoff sensitivity is reduced. Therefore, the learning dynamics is generally more stable for $\kappa<1$, and convergence to a fixed point in the center of the strategy space occurs for a larger set of parameter combinations. All the analysis of the baseline scenario still applies.

In the belief learning case ($\kappa=0$) the rescaled payoff sensitivity is $\tilde{\beta}=\beta\alpha$. This means that the coordinates of the fixed points do not depend on $\alpha$, see Eqs.\eqref{eq:fixedpoints-newcoordinates2} (as in Figure \ref{fig:summaryfigure}). One can show that the fixed points then correspond to the Quantal Response Equilibria (QRE) of the game. QRE were introduced by \cite{mckelvey1995quantal} to allow for boundedly rational players, in particular to include the possibility that players make errors. Here the QRE $x^\star$ and $y^\star$ are given by the solutions to 
\begin{equation}
\begin{split}
\overline{\Pi_2^R}(y^\star)-\overline{\Pi_1^R}(y^\star) = \frac{1}{\beta}\ln\frac{1-x^\star}{x^\star}, \\
\overline{\Pi_2^C}(x^\star)-\overline{\Pi_1^C}(x^\star) = \frac{1}{\beta}\ln\frac{1-y^\star}{y^\star}.
\end{split}
\label{eq:belieflearningalpha1}
\end{equation}
For small values of $\beta$ the QRE are in the center of the strategy space, whereas increasing values of $\beta$ bring the QRE closer to the NE. In the limit $\beta\rightarrow\infty$, the QRE coincide with the NE. 

With $\kappa=0$ the stability condition is (in Matching Pennies games)
\begin{equation}
\frac{\beta\alpha}{\sqrt{2\alpha-\alpha^2}}\left| A \right|\leq 1.
\label{eq:asymmetricstability-av}
\end{equation}

Differently from Eq. \eqref{eq:asymmetricstability-2}, the derivative of the left hand side  with respect to $\alpha$ is positive and so longer memory promotes stability. For general $\kappa$, the numerator in Eq. \eqref{eq:asymmetricstability-av} is $\beta\left[1-(1-\alpha)(1-\kappa) \right]$, so the derivative is positive when $\alpha>\kappa$. The effect of memory on stability is thus not trivial: in the belief learning limit, long memory promotes stability, but when $\alpha<\kappa$ long memory promotes instability. To the best of our knowledge, we are the first to identify the role of memory on instability in this class of learning rules.

In the limit $\alpha\rightarrow 0$, the left hand side of Eq. \eqref{eq:asymmetricstability-av} goes to zero, so stability is ensured for all parameter values. For $\beta=+\infty$, we recover the well known result of \cite{miyazawa1961convergence} and \cite{monderer1996a2}, namely that in non-degenerate $2\times 2$ games fictitious play would converge to the NE. For other values of $\beta$, we recover the results of \cite{fudenberg1993learning} and \cite{benaim1999mixed}, namely that in $2\times 2$ games stochastic fictitious play would converge to the QRE.

\subsection{Reinforcement learning}
\label{sec:reinforcement}

We now relax the constraint $\delta=1$, and allow the players to give different weight to the actions that were and were not taken. For analytical tractability we assume that the players have perfect memory, $\alpha=0$. We also assume $\kappa>0$. (With $\alpha=0$, $\beta$ does not determine the existence and properties of the fixed points, as in Figure \ref{fig:summaryfigure}; so we could just set $\tilde{\beta}=\beta\left[1-(1-\alpha)(1-\kappa) \right]=1$.) As we cannot use the coordinate transformation \eqref{eq:coordinatetransformation}, we obtain the fixed points directly from Eq. \eqref{eq:EWA3}.

By replacing the parameter restrictions in Eq. \eqref{eq:EWA3}, it is possible to show that there are now potentially ten fixed points for a given value of $\delta$. We give the expressions of all fixed points explicitly or implicitly in Appendix \ref{sec:reinfapp}. Four fixed points are the pure strategy profiles, with $(x,y)$ equal to $(0,0)$, $(0,1)$, $(1,0)$ and $(1,1)$. In four additional fixed points either $x$ or $y$ are 0 or 1, but not both, i.e. these fixed points are of the form $(0,y_1)$, $(1,y_2)$, $(x_1,0)$ and $(x_2,1)$. Finally, two fixed points have both $x$ and $y$ different from 0 and 1, i.e. $(x_3,y_3)$ and $(x_4,y_4)$. Only the pure strategy profiles are fixed points for all choices of model parameters; the other fixed points may or may not exist, depending on the choice of $\delta$ or of the payoffs. 

In terms of stability, for each fixed point corresponding to the pure strategy profiles $(x,y)=\{(0,0),(0,1),(1,0),(1,1)\}$, we specify  the two eigenvalues of the Jacobian at that fixed point:
\begin{equation}
\begin{split}
(x,y)=(0,0)=(s_2^R,s_2^C)\hspace{20pt}\rightarrow\hspace{20pt} \left(e^{\beta(b\delta-d)},e^{\beta(f\delta-h)}\right), \\
(x,y)=(0,1)=(s_2^R,s_1^C)\hspace{20pt}\rightarrow\hspace{20pt} \left(e^{\beta(a\delta-c)},e^{\beta(h\delta-f)}\right), \\
(x,y)=(1,0)=(s_1^R,s_2^C)\hspace{20pt}\rightarrow\hspace{20pt} \left(e^{\beta(d\delta-b)},e^{\beta(e\delta-g)}\right), \\
(x,y)=(1,1)=(s_1^R,s_1^C)\hspace{20pt}\rightarrow\hspace{20pt} \left(e^{\beta(c\delta-a)},e^{\beta(g\delta-e)}\right). 
\end{split}
\label{eq:eigdelta}
\end{equation}

If we set $\delta=1$, we get the same result of Proposition \ref{prop1} in Section \ref{sec:purestrat}, namely that only the pure strategy NE are stable. However, by taking $\delta<1$ it is also possible to make the other pure strategy profiles potentially stable, by effectively reducing the ``perceived'' value of the payoffs at the NE (i.e., the players do not realize they could earn higher payoff if they unilaterally switched).  We explain this with an example. Consider the action profile $(s_1^R,s_1^C)$, and assume that $(s_2^R,s_1^C)$ is a NE. This means that $c>a$, and so from Eq. \eqref{eq:eigdelta} the first eigenvalue of $(x,y)=(1,1)$ is greater than one for $\delta=1$. So the pure strategy profile $(s_1^R,s_1^C)$ is unstable. But if $\delta\neq 1$, the condition for the fixed point to be unstable becomes $c\delta - a > 0$, i.e. $c>a/\delta$. Therefore, provided $a>0$, for the NE to be the unique stable fixed point of EWA, the payoffs at the NE must be larger by a factor $1/\delta$ than the payoffs at $(s_2^R,s_1^C)$. In the other cases, non-NE can also be stable fixed points. Mathematically, this means that for $\delta<1$ the dynamics can be stuck in local minima that are hard to justify in terms of rationality, as each player could potentially improve her payoff by switching action. However, as the players do not consider forgone payoffs, they cannot realize this, and keep playing the same action. 

\begin{figure}[h]
\centering
\includegraphics[width=1\textwidth]{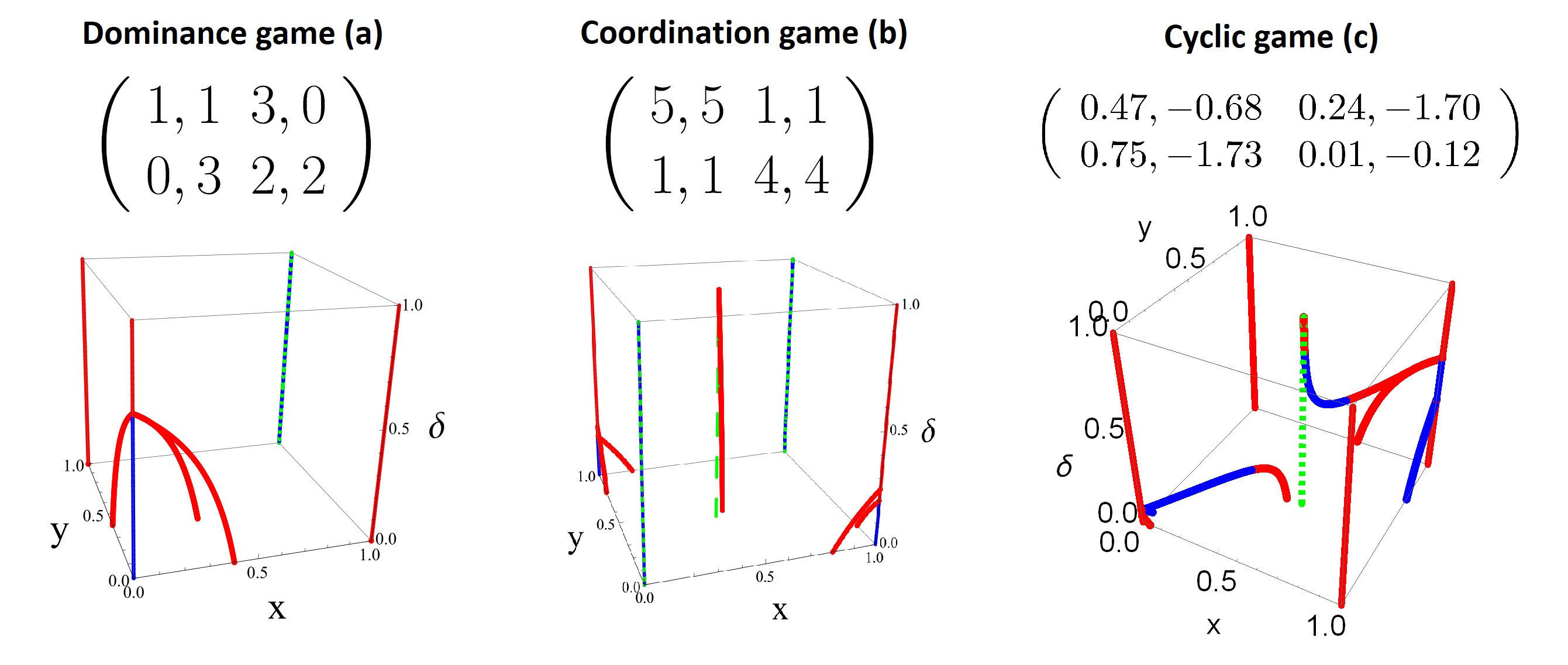}
\caption{Bifurcation diagram showing the fixed points $(x,y)$ as the $\delta$ parameter is varied between 0 and 1. Blue (red) lines represent stable (unstable) fixed points. Dashed green lines represent NE. Lower values of $\delta$ increase the likelihood to have stable fixed points that do not coincide with the NE. }
\label{fig:calculation_delta_cube}
\end{figure}

Explicit examples are given in Figure \ref{fig:calculation_delta_cube}. The axes $x$ and $y$ give the positions of the fixed points for a specific value of $\delta$, and the vertical axis $\delta$ shows how $x$ and $y$ vary with this parameter. When the lines are blue it means that the fixed point is stable, and when red the fixed point is unstable. Dashed green lines represent NE. When a NE coincides with a fixed point the lines are shown blue or red with green dashes.

In panel (a) the game is dominance-solvable, with a unique NE at $(x,y)=(1,1)$ (it is a Prisoner Dilemma). This NE is stable for all values of $\delta$, but the Pareto-optimal pure strategy profile $(x,y)=(0,0)$ is also stable for $\delta \in [0,2/3]$.  The other solutions of the type $(0,y)$ or $(x,0)$ --- on the faces of the cube --- or $(x,y)$, are always unstable. The situation is similar in panel (b), except that the payoff matrix describes a coordination game with two pure strategy NE. The other two pure strategy profiles are stable for $\delta<1/5$ or $\delta<1/4$, as can be calculated from Eq. \eqref{eq:eigdelta}.  

Finally, case (c) is a cyclic game with the maximal number of fixed points, as all solutions exist. When $\delta=0$ both fixed points $(0,y_1)$ and $(1,y_2)$ are stable; as $\delta$ is increased, the solution of the type $(x_3,y_3)$ or $(x_4,y_4)$ with $x,y<0.5$ becomes stable. As $\delta$ is further increased the pure strategy profile $(1,1)$ is stable, and finally it is the solution of the type $(x_3,y_3)$ or $(x_4,y_4)$ with $x,y>0.5$ that becomes stable. For $\delta>0.82$ all solutions are unstable, and the learning dynamics does not converge to any fixed point. Note that in this game no stable fixed point corresponds to the NE, and can be arbitrarily far.

\subsection{Stochastic learning}
\label{sec:stoch}

When playing a game, except for very specific experimental arrangements \citep{conlisk1993adaptation}, real players update their strategies after observing a single action by their opponent, and so they do not not know her mixed strategy vector. This questions whether the analysis of the deterministic dynamics so far provides robust conclusions. In this section we provide some simulations arguing that it does.\footnote{It is beyond the scope of this paper to systematically study the effect of noise on the learning dynamics. We refer the reader to \cite{galla2009intrinsic} for a study on the effect of noise on learning, and to \cite{crutchfield1982fluctuations} for a more general discussion on the effect of noise on dynamical systems. }

\begin{figure}[h]
\centering
\includegraphics[width=0.7\textwidth]{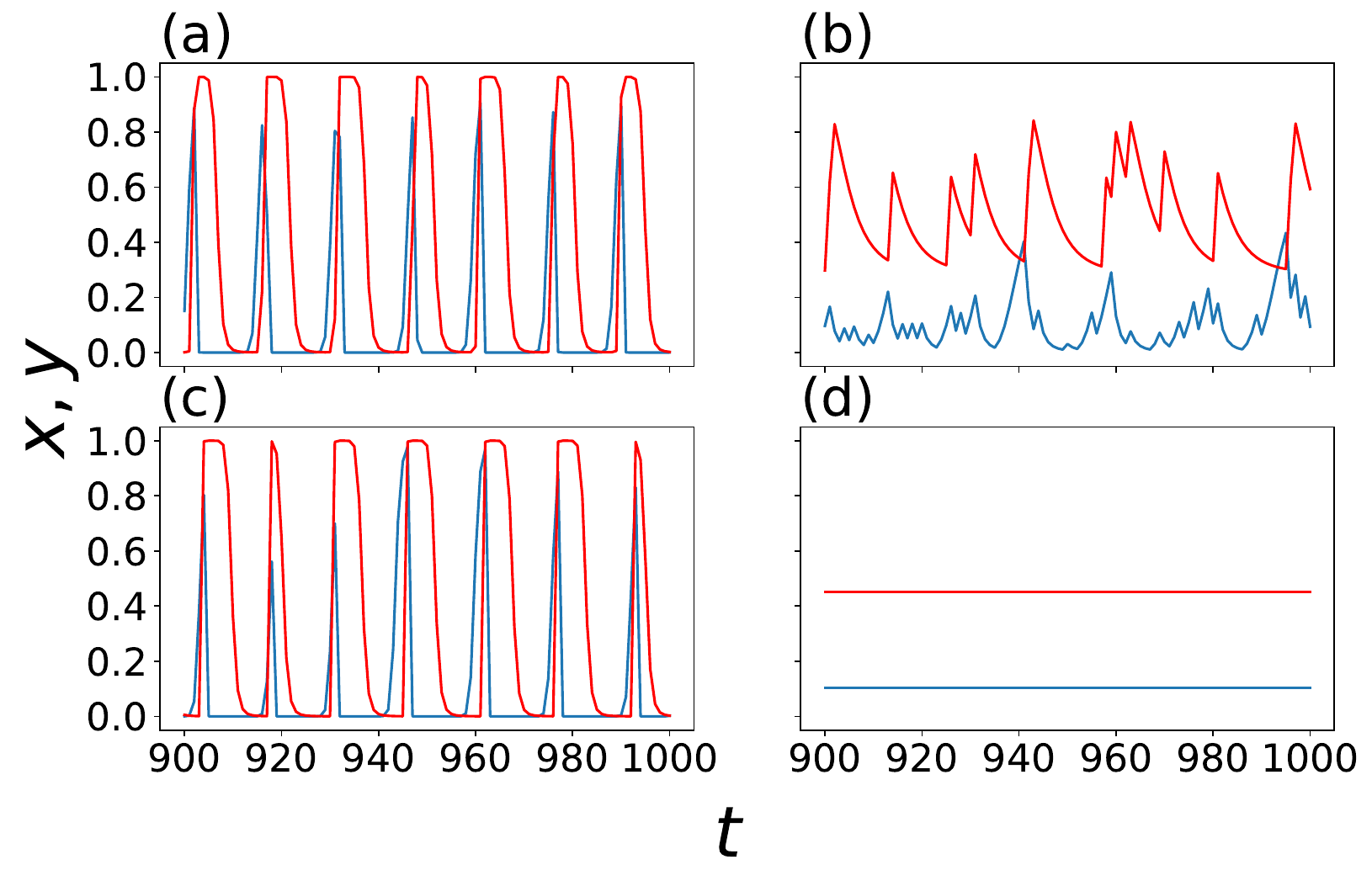}
\caption{Time series of the probabilities $x$ (in blue) and $y$ (in red) of the learning dynamics in a cyclic game. Top panels represent stochastic learning, bottom panels the corresponding deterministic learning. In all cases the payoff combinations are $A=-C=-3.4$ and $B=-D=-2.5$, and the memory loss is $\alpha=0.2$. In panel (c) the deterministic dynamics converges to a chaotic attractor ($\beta=1$), while in panel (d) it reaches a fixed point ($\beta=0.1$). }
\label{fig:TimeSeries_stochastic_limitcycle}
\end{figure}

When the deterministic dynamics moves close to the boundaries of the strategy space, we expect that the corresponding stochastic dynamics behaves similarly, with some occasional fluctuation. This is because the probability to play a different action than the one being played at the border of the strategy space is very small. If instead there is a unique stable fixed point in the center of the strategy space, we expect the fluctuations to be substantial, as any action can be chosen roughly with equal probability. 

In Figure \ref{fig:TimeSeries_stochastic_limitcycle} we report examples that confirm this intuition. In panels (a) and (c) there are no stable fixed points, and the deterministic dynamics follows a chaotic attractor where players play mixed strategies at the border of the parameter space (panel (c)). The corresponding stochastic dynamics is very similar (in fact, we show in Supplementary Appendix \ref{sec:suppstochlearning} that the stochastic dynamics is also chaotic). The situation is very different in panels (b) and (d). Here, the deterministic dynamics converges to a fixed point in the center of the strategy space (d), while the stochastic version substantially fluctuates around that fixed point.

\section{Conclusion}
\label{sec:conclusion}

In this paper we have followed the literature that assumes boundedly rational players engaging in an infinitely repeated game and updating their stage-game strategies using an adaptive learning rule, here Experience-Weighted Attraction (EWA). We have characterized the asymptotic outcome of this learning process in $2\times 2$ games, classifying when it would converge to a Nash Equilibrium (NE), when it would converge to a different fixed point, or when it would follow limit cycles of chaos. 

Most of the works in the literature focus on the convergence properties of one or two learning rules. As EWA generalizes several learning rules that have been extensively studied in the literature --- reinforcement learning, various forms of fictitious play, best response dynamics and also replicator dynamics with finite memory --- our contribution is to provide a systematic characterization, or taxonomy, of learning dynamics, extending results that are only valid for extreme parameterizations of EWA, showing new phenomena. These include instability of Pareto-inefficient NE, stability of fixed points of mutual cooperation, and an ambiguous effect of memory on stability.  Our taxonomy is also useful to provide theoretical guidance on the learning dynamics to be expected in experiments, as EWA is widely used to model learning behavior in several classes of games.

\bibliographystyle{C:/Users/marco/Dropbox/Latex/econ}
\bibliography{./refs/ref}

\newpage
\appendix
\renewcommand\thefigure{\thesection.\arabic{figure}} 
\counterwithin{theorem}{section}

\section{Proof of Proposition \ref{prop1}}
\label{sec:proofprop1}

In order to study the properties of the pure strategy NE we need to consider the learning dynamics in the original coordinates (the pure strategies map into infinite elements in the transformed coordinates). The EWA dynamics reads (using \eqref{eq:EWA3} with $\delta=1$, $\kappa=1$ and the payoff matrix \eqref{eq:payoff}):
\begin{equation}
\begin{split}
x(t+1) = \frac{x(t)^{1-\alpha}e^{\beta \left( a y(t) + b (1-y(t) \right) }}{x(t)^{1-\alpha}e^{\beta \left( a y(t) + b (1-y(t) \right) } + (1-x(t))^{1-\alpha}e^{\beta \left( c y(t) + d (1-y(t) \right) }}, \\
y(t+1) = \frac{y(t)^{1-\alpha}e^{\beta \left( e x(t) + f (1-x(t) \right) }}{y(t)^{1-\alpha}e^{\beta \left( e x(t) + f (1-x(t) \right) }  + (1-y(t))^{1-\alpha}e^{\beta \left( g x(t) + h (1-x(t) \right) }}.
\end{split}
\label{eq:EWA-originalcoordinates}	
\end{equation}

From Eq. \eqref{eq:EWA-originalcoordinates} we can see that the pure strategies $(x,y) \in \{ (0,0),(0,1),(1,0),(1,1) \}$ are all fixed points of the dynamics. Let us study their stability properties. We get a Jacobian

\begin{equation}
J=\begin{pmatrix}
  J_{11} & J_{12} \\
  J_{21} & J_{22}
 \end{pmatrix},
\label{eq:jacobian}	
\end{equation}
with
\begin{equation}
\begin{split}
J_{11} = \frac{(1 - \alpha) (x - x^2)^{\alpha } e^{\beta  (y (a-b-c+d)+b-d)}}{\left(x (1-x)^{\alpha } e^{\beta  (y (a-b-c+d)+b-d)}-(x-1) x^{\alpha }\right)^2}, \\
J_{12} = \frac{\beta  (x-x^2)^{\alpha +1} (a-b-c+d) e^{\beta  (y (a-b-c+d)+b-d)}}{\left(x (1-x)^{\alpha } e^{\beta  (y (a-b-c+d)+b-d)}-(x-1) x^{\alpha }\right)^2}, \\
J_{21} = \frac{\beta (y-y^2)^{\alpha +1} (e-f-g+h) e^{\beta  (x (e-f-g+h)+f-h)}}{\left(y (1-y)^{\alpha } e^{\beta  (x (e-f-g+h)+f-h)}-(y-1) y^{\alpha }\right)^2}, \\
J_{22} = \frac{(1-\alpha ) (y-y^2)^{\alpha } e^{\beta  (x (e-f-g+h)+f-h)}}{\left(y (1-y)^{\alpha } e^{\beta  (x (e-f-g+h)+f-h)}-(y-1) y^{\alpha }\right)^2}.
\end{split}
\label{eq:jacobian-elements}	
\end{equation}

As it can be seen by taking the appropriate limits in Eqs. \eqref{eq:jacobian-elements}, for all profiles of pure strategies the Jacobian has infinite elements along the main diagonal and null elements along the antidiagonal. This means that the profiles of pure strategies --- and in particular the pure strategy NE --- are ``infinitely'' unstable.  

However, when $\alpha = 0$ the pure strategy NE become stable. Consider the profile of pure strategies in which both players choose action $s_1$. This corresponds to $x=y=1$, and gives a Jacobian 
\begin{equation}
J=\begin{pmatrix}
  e^{-\beta(a-c)} & 0 \\
  0 & e^{-\beta(e-g)}
 \end{pmatrix}.
\end{equation}
The eigenvalues can be seen on the main diagonal, and are stable when $a>c$ and $e>g$. Under these conditions $(s_1^R,s_1^C)$ is a pure strategy NE. The argument is similar for all other pure strategy profiles. 

\section{Proof of Proposition \ref{prop2}}
\label{sec:proofprop2}

We first consider claim \textit{(i)}. Since $B=0$, there is always a fixed point $(\tilde{x}^\star,\tilde{y}^\star) = (0,0)$. This fixed point is stable if (from Eq. \eqref{eq:symmetricstability-3})

\begin{equation}
\frac{\beta}{\alpha}\left| A \right| \leq 1.
\label{eq:symmetricstability-4}
\end{equation}
So, as long as $\tilde{x}^\star = 0$ is the unique fixed point, it is stable. 

We then consider claim \textit{(ii)}, and in particular the lower bound, $ \frac{\beta}{\alpha}\left| A \right| \rightarrow 1^+$. Further to the central fixed point, there are two lateral fixed points $\tilde{x}^\star = \pm \epsilon$, where $\epsilon$ is an arbitrarily small number. Thanks to the symmetry of the game, we focus on a profile of mixed strategies given by $(\tilde{x}^\star,\tilde{x}^\star)$. (A similar argument is valid for fixed points of the type $(\tilde{x}^\star,-\tilde{x}^\star)$.)  To second order, $\cosh \tilde{x}^\star \approx 1 + \left(\tilde{x}^\star\right)^2/2$. The stability condition becomes
\begin{equation}
\frac{\alpha}{\beta} \left(1 + \frac{\left(\tilde{x}^\star\right)^2}{2}\right) \left(1 + \frac{\left(\tilde{x}^\star\right)^2}{2}\right) - \left| A \right| \geq 0,
\label{eq:symmetricstability-5}
\end{equation}
i.e. 
\begin{equation}
\left(\tilde{x}^\star\right)^2 \geq \frac{\beta}{\alpha}\left| A \right| - 1.
\label{eq:symmetricstability-6}
\end{equation}
Now, we Taylor expand $\Psi(\tilde{x}^\star)$ (defined in Section \ref{sec:symmetric}) to third order (first order would just yield $\tilde{x}^\star=0$)) and solve $\tilde{x}^\star=\Psi(\tilde{x}^\star)$. Apart from the null solution, we get
\begin{equation}
\left(\tilde{x}^\star\right)^2 = \frac{3\left( \frac{\beta^2 A^2}{\alpha^2}-1 \right)}{ \frac{\beta^2 A^2}{\alpha^2} \left( 1 +  \frac{\beta^2 A^2}{\alpha^2} \right)}.
\label{eq:symmetricstability-7}
\end{equation}
It is easily checked that for $ \frac{\beta}{\alpha}\left| A \right| \rightarrow 1^+$, the condition \eqref{eq:symmetricstability-6} is satisfied: the  fixed points whose components are the ``lateral solutions'' are stable. Therefore, there is a supercritical pitchfork bifurcation at the value $ \frac{\beta}{\alpha}\left| A \right| = 1$.

The upper bound, namely $ \frac{\beta}{\alpha}\left| A \right| \rightarrow \infty$, is easily dealt with. As discussed in Section \ref{sec:symmetric}, in this limit the fixed point $\tilde{x}^\star$ is given by $\tilde{x}^\star \approx \pm \frac{\beta}{\alpha}\left| A \right|$. Now, for $ \frac{\beta}{\alpha}\left| A \right| \rightarrow + \infty$ the hyperbolic cosine can be approximated by
\begin{equation}
\cosh \left( \frac{\beta}{\alpha}\left| A \right| \right) \approx \exp \left( \frac{\beta}{\alpha}\left| A \right| \right) / 2.
\label{eq:symmetricstability-8}
\end{equation}
We can rewrite the stability condition as:
\begin{equation}
\frac{4\beta}{\alpha}\left| A \right| \exp \left( -2\frac{\beta}{\alpha}\left| A \right| \right) \leq 1.
\label{eq:symmetricstability-9}
\end{equation}
For $ \frac{\beta}{\alpha}\left| A \right| \rightarrow \infty$, the left hand side of the above equation goes to zero, so the inequality obviously holds.

Finally, the proof of \textit{(iii)} is identical to the proof of the upper bound for $\frac{\beta}{\alpha}\left| A \right|$, in that the same arguments apply to sufficiently large values of $\frac{\beta}{\alpha}\left| B \right|$ (provided that $B\gg A$). 

\section{Fixed points of reinforcement learning}
\label{sec:reinfapp}

The fixed points are obtained setting $x(t+1)=x(t)=x^\star$ and $y(t+1)=y(t)=y^\star$ in Eq. \eqref{eq:EWA3}, with $\alpha=0$ and $\kappa=1$. This gives 
\begin{equation}
\begin{split}
x^\star = \frac{x^\star e^{\beta[\delta+(1-\delta)x^\star](ay^\star+b(1-y^\star))}}{x^\star e^{\beta[\delta+(1-\delta)x^\star](ay^\star+b(1-y^\star))}+(1-x^\star) e^{\beta[\delta+(1-\delta)(1-x^\star)](cy^\star+d(1-y^\star))}},\\
y^\star = \frac{y^\star e^{\beta[\delta+(1-\delta)y^\star](ex^\star+f(1-x^\star))}}{y^\star e^{\beta[\delta+(1-\delta)y^\star](ex^\star+f(1-x^\star))}+(1-y^\star) e^{\beta[\delta+(1-\delta)(1-y^\star)](gx^\star+h(1-x^\star))}}.
\end{split}
\label{eq:reinfapp1}
\end{equation}

It is easily checked that all pure strategy profiles are fixed points. Four additional solutions can be found noticing that when either $x^\star$ or $y^\star$ are 0 or 1, the respective equation holds as an identity. It is then possible to find the other solution by solving the other equation. These give fixed points
\begin{equation}
\begin{split}
(0,y_1) = \left(0,\frac{(1-\delta)h+\delta(h-f)}{(1-\delta)(f+h)}\right), \\
(1,y_2)= \left(1,\frac{(1-\delta)g+\delta(g-e)}{(1-\delta)(g+e)}\right), \\
(x_1,0)= \left(\frac{(1-\delta)d+\delta(d-b)}{(1-\delta)(d+b)},0\right), \\
(x_2,1)= \left(\frac{(1-\delta)c+\delta(c-a)}{(1-\delta)(c+a)},1\right). 
\end{split}
\label{eq:reinfapp2}
\end{equation}
Of course for these solutions to exist it has to be $0<y_1,y_2,x_1,x_2<1$. The two final solutions $(x_3,y_3)$ and $(x_4,y_4)$ are obtained when the arguments of the exponentials in Eq. \eqref{eq:reinfapp1} are identical, i.e. when
\begin{equation}
\begin{split}
(\delta+(1-\delta)x^\star)(ay^\star+b(1-y^\star))=(\delta+(1-\delta)(1-x^\star))(cy^\star+d(1-y^\star)),\\
(\delta+(1-\delta)y^\star)(ex^\star+f(1-x^\star))=(\delta+(1-\delta)(1-y^\star))(gx^\star+h(1-x^\star)).
\end{split}
\label{eq:reinfapp3}
\end{equation}
The expression for these solutions is very complicated and uninsightful. We only report this expression for $\delta=0$ and symmetric games, 
\begin{equation}
x^\star=y^\star=\frac{b-c+2d\pm\sqrt{b^2-2bc+c^2+4ad}}{2(b-a+d-c)}.
\end{equation}
We do not report the eigenvalues of other fixed points than the pure strategy profiles (Eq. \eqref{eq:eigdelta}) because their expression is complicated and uninsightful.

\makeatletter
\renewcommand \thesection{S\@arabic\c@section}
\renewcommand\thetable{S\@arabic\c@table}
\renewcommand \thefigure{S\@arabic\c@figure}
\makeatother
\setcounter{section}{0}
\setcounter{figure}{0}

\newpage

\begin{center}
{\Huge \textbf{Supplementary Appendix}}
\end{center}

\section{Replicator dynamics with finite memory}
\label{sec:repldyn}

Here we show that the EWA equations \eqref{eq:EWA3} have a continuous time limit that corresponds to a generalized version of replicator dynamics having finite memory, instead of infinite memory as in the standard case. We present an alternative derivation with respect to previous papers. \cite{sato2003coupled} assume that the evolution of the attractions takes place at a different timescale than the evolution of the probabilities, and Galla and Farmer (2013, Supplementary Information, Section II) use a Lagrange multiplier method. 

Here we simply start from Eq. \eqref{eq:EWA3} and take the limit $\alpha \rightarrow 0$, $\beta \rightarrow 0$, such that the ratio $\alpha/\beta$ is finite. In this limit $\tilde{\kappa}=\kappa$, and we set $\kappa=1$ without loss of generality.\footnote{As the case $\kappa=0$ is excluded from the condition on the steady state of the experience, $\kappa$ is just a constant that multiplies $\beta$.} We also only perform the calculations for $x(t)$, as the calculations for $y(t)$ are identical. For notational simplicity, we denote here $x(t)$ by $x_t$, and $y(t)$ by $y_t$. Taking logs in Eq. \eqref{eq:EWA3} we get
\begin{multline}
\ln x_{t+1} = (1-\alpha)\ln x_t + \beta [\delta+(1-\delta)x_t]\overline{\Pi_1^R}(y_t)-\\
 \ln\left( x_t^{1-\alpha}e^{\beta[\delta+(1-\delta)x_t]\overline{\Pi_1^R}(y_t)} + (1-x_t)^{1-\alpha}e^{\beta[\delta+(1-\delta)(1-x_t)]\overline{\Pi_2^R}(y_t)} \right).
\label{eq:repl1}
\end{multline}
The logarithm of the denominator can be greatly simplified by taking the limit $\alpha \rightarrow 0$, $\beta \rightarrow 0$. In this limit
\begin{equation}
x_t^{1-\alpha}=x_t x_t^{-\alpha}=x_t e^{\ln x_t^{-\alpha}} = x_t e^{-\alpha \ln x_t} \approx x_t (1-\alpha \ln x_t)
\end{equation}
and
\begin{equation}
e^{\beta[\delta+(1-\delta)x_t]\overline{\Pi_1^R}(y_t)} \approx \left(1+\beta[\delta+(1-\delta)x_t]\overline{\Pi_1^R}(y_t)\right).
\end{equation}
By ignoring terms of $O(\alpha^2)$ (or equivalently $O(\beta^2)$, as the ratio $\alpha/\beta$ is finite) we can then write 
\begin{multline}
\ln\left( x_t^{1-\alpha}e^{\beta[\delta+(1-\delta)x_t]\overline{\Pi_1^R}(y_t)} + (1-x_t)^{1-\alpha}e^{\beta[\delta+(1-\delta)(1-x_t)]\overline{\Pi_2^R}(y_t)} \right) \approx \\
\ln\left( x_t\left(1-\alpha\ln x_t + \beta[\delta+(1-\delta)x_t]\overline{\Pi_1^R}(y_t)\right)+\right. \\
\left. (1-x_t)\left(1-\alpha\ln (1-x_t) + \beta[\delta+(1-\delta)(1-x_t)]\overline{\Pi_2^R}(y_t)\right) \right) =\\
\ln \left(1+x_t\left(-\alpha\ln x_t + \beta[\delta+(1-\delta)x_t]\overline{\Pi_1^R}(y_t)\right)+\right.\\
\left.(1-x_t)\left(-\alpha\ln (1-x_t) + \beta[\delta+(1-\delta)(1-x_t)]\overline{\Pi_2^R}(y_t)\right)\right)\approx\\
x_t\left(-\alpha\ln x_t + \beta[\delta+(1-\delta)x_t]\overline{\Pi_1^R}(y_t)\right)+(1-x_t)\left(-\alpha\ln (1-x_t) + \beta[\delta+(1-\delta)(1-x_t)]\overline{\Pi_2^R}(y_t)\right).
\end{multline}
Replacing this in Eq. \eqref{eq:repl1} and rearranging gives
\begin{multline}
\ln x_{t+1}-\ln x_t = \beta \left( [\delta+(1-\delta)x_t]\overline{\Pi_1^R}(y_t)-x_t [\delta+(1-\delta)x_t]\overline{\Pi_1^R}(y_t) - \right. \\
\left.(1-x_t)[\delta+(1-\delta)(1-x_t)]\overline{\Pi_2^R}(y_t) \right)-\alpha \left( \ln x_t-x_t \ln x_t - (1-x_t)\ln (1-x_t) \right).
\end{multline}
It is possible to divide everything by $\beta$ and rescale time so that one time unit is given by $\beta$. Then in the limit $\beta \rightarrow 0$ the left hand side of the above equation is 
\begin{equation}
\lim_{\beta \rightarrow 0} \frac{\ln x_{t+\beta}-\ln x_t}{\beta}=\frac{\dot{x}}{x}
\end{equation}
and the learning dynamics for $x$ can be written in continuous time as
\begin{multline}
\frac{\dot{x}}{x} = [\delta+(1-\delta)x]\overline{\Pi_1^R}(y)-x [\delta+(1-\delta)x]\overline{\Pi_1^R}(y) -  \\
(1-x)[\delta+(1-\delta)(1-x)]\overline{\Pi_2^R}(y) -\frac{\alpha}{\beta} \left( \ln x-x \ln x - (1-x)\ln (1-x) \right).
\label{eq:repl2}
\end{multline}
This is in general the continuous time approximation of the EWA dynamics in Eq. \eqref{eq:EWA3}. In the case $\delta=1$, replacing the expressions for $\overline{\Pi_1^R}$ and $\overline{\Pi_2^R}$, we get
\begin{equation}
\frac{\dot{x}}{x}=ay+b(1-y)-(axy+bx(1-y)+c(1-x)y+d(1-x)(1-y))-\frac{\alpha}{\beta} \left(\ln x - H(x)\right),
\label{eq:repl3}
\end{equation}
where $H(x)=x \ln x + (1-x)\ln (1-x)$ is the information entropy of mixed strategy $(x,1-x)$. This is the dynamics analyzed in \cite{sato2003coupled}. If $\alpha=0$, i.e. with infinite memory, the above equation reduces to the standard form of two-population replicator dynamics \citep{hofbauer1998evolutionary}. 

It is useful to analyze the stability of Eq. \eqref{eq:repl3} in cyclic games. We rewrite the replicator dynamics \eqref{eq:repl3} in terms of $A$, $B$, $C$ and $D$, factor a $1-x$ term and write the corresponding equation for $y$:
\begin{equation}
\begin{split}
\dot{x} = x(1-x)\left(4Ay+2(B-A)+\frac{\alpha}{\beta}(\ln (1-x) - \ln x)\right)  , \\
\dot{y} = y(1-y)\left(4Cx+2(D-C)+\frac{\alpha}{\beta}(\ln (1-y) - \ln y)\right)  .
\end{split}
\end{equation}
In line with the analysis in Section \ref{sec:asymmetric}, we focus on the specific case in which $B=D=0$ and $C=-A$, i.e. Matching Pennies. In this case the fixed points of the replicator dynamics are $(0,0)$, $(0,1)$, $(1,0)$, $(1,1)$ and $(1/2,1/2)$. The fixed points $(0,0)$, $(0,1)$, $(1,0)$ and $(1,1)$ are always unstable, for any value of $\alpha$. The eigenvalues for the fixed point $(1/2,1/2)$ are 
\begin{equation}
\lambda_\pm=-\frac{\alpha}{\beta}\pm iA.
\end{equation}
As the stability of a fixed point of a continuous dynamical system is determined by whether the real part of the eigenvalues is positive, it is easy to see that $(1/2,1/2)$ is always stable for $\alpha>0$. Therefore, the replicator dynamics with finite memory always converges to the mixed strategy NE. When $\alpha=0$ the fixed point becomes marginally stable, and the learning dynamics circles around the NE. This recovers a standard result in evolutionary game theory \citep{hofbauer1998evolutionary}. Note however that if $B\neq 0$ or $D\neq 0$ the position of the fixed point in the strategy space becomes dependent on $\alpha/\beta$, and can be arbitrarily far from the mixed strategy NE.

\section{Additional results}
\label{sec:additionalresults}

\subsection{Proof of proposition \ref{prop0}}
\label{sec:proofprop0}

\begin{proof}

We only prove that we have a coordination game (defined by $a>c$, $e>g$, $d>b$, $h>f$) if and only if $|A|>|B|$, $|C|>|D|$, $A>0$, $C>0$. The other cases are similar. 

We first prove that a coordination game implies $|A|>|B|$, $|C|>|D|$, $A>0$ and $C>0$. First of all, when $a>c$ and $d>b$, $A$ is positive, $A = \frac{1}{4} (a-c+d-b)>0$. Then, because $A>0$, the expression $|A|>|B|$ can be written as $A>|B|$. If $B>0$, this expression can further be written as $A-B>0$. This inequality is indeed satisfied from the condition $d-b>0$ that defines a coordination game, i.e. $A-B = 2(d-b)>0$. If $B<0$, we need to check that $A+B>0$, and this is obtained from the other coordination game condition $a-c>0$, i.e. $A+B=2(a-c)>0$. The argument for $C$ and $D$ is analogous.

We next consider the viceversa---that the conditions on $A,B,C$ and $D$ imply a coordination game. Consider $B>0$ without loss of generality. Because also $A>0$, we can remove the absolute values in the condition $|A|>|B|$, which becomes $A-B>0$. This implies $d>b$. We still need to show that $a>c$, which does not simply follow from the definition of $A$, $A=\frac{1}{4}(a-c+d-b)>0$. Indeed, the definition of $A$ only implies $a-c>-(d-b)$, but because we just proved that $d-b>0$, this condition could also be satisfied with $a-c<0$. However, if $a-c<0$, we have $B=\frac{1}{4}(a-c-(d-b))<0$, which is in contradiction with our former assumption. The same considerations apply to $C$ and $D$.
\end{proof}

\subsection{Chaotic dynamics}
\label{sec:chaoticdynamics}

\begin{figure}[h]
\centering
\includegraphics[width=0.85\textwidth]{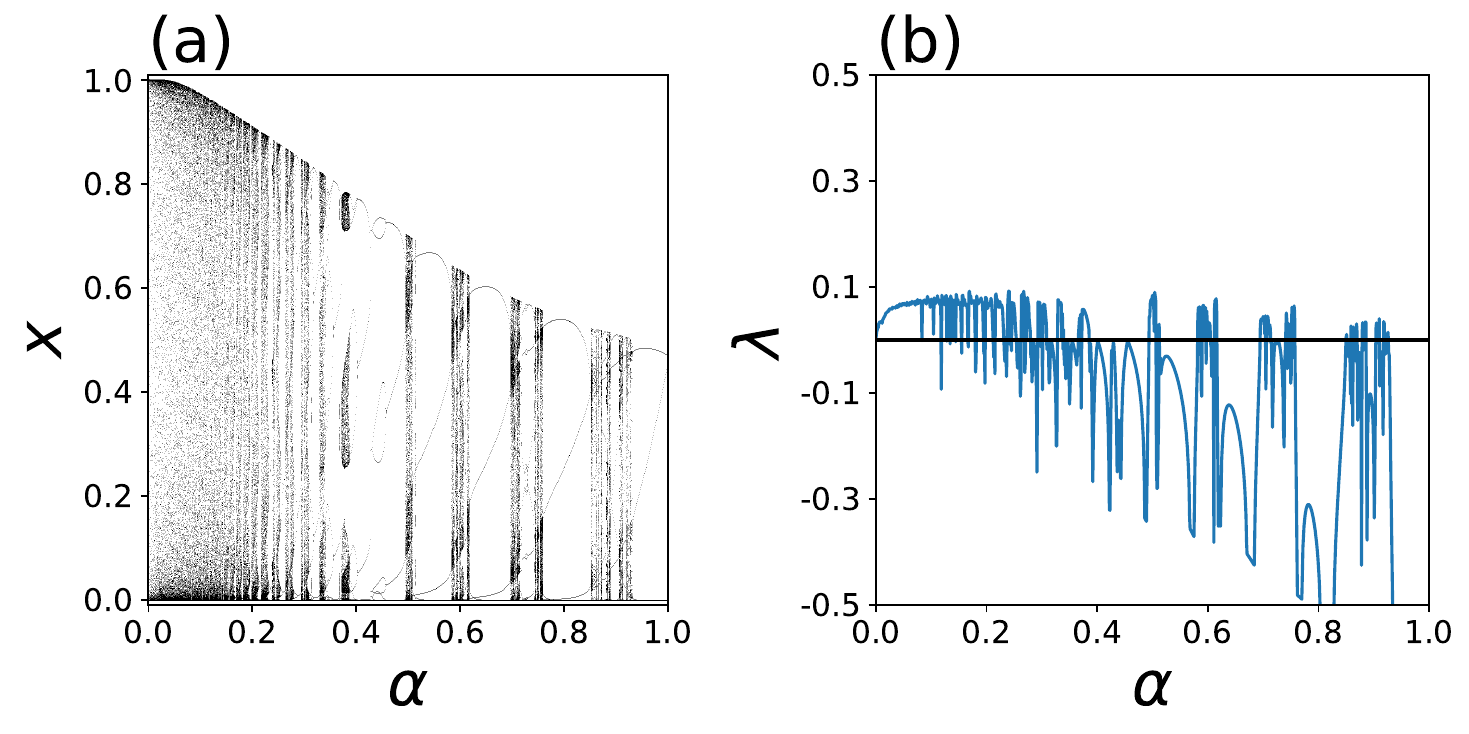}
\caption{Bifurcation diagram and largest Lyapunov exponent $\lambda$ as $\alpha$ is varied between 0 and 1. Cyclical and chaotic dynamics alternate, with chaos being more likely for small values of $\alpha$.}
\label{fig:plot_lyapunov_exponents_alpha}
\end{figure}

\begin{figure}[h]
\centering
\includegraphics[width=0.9\textwidth]{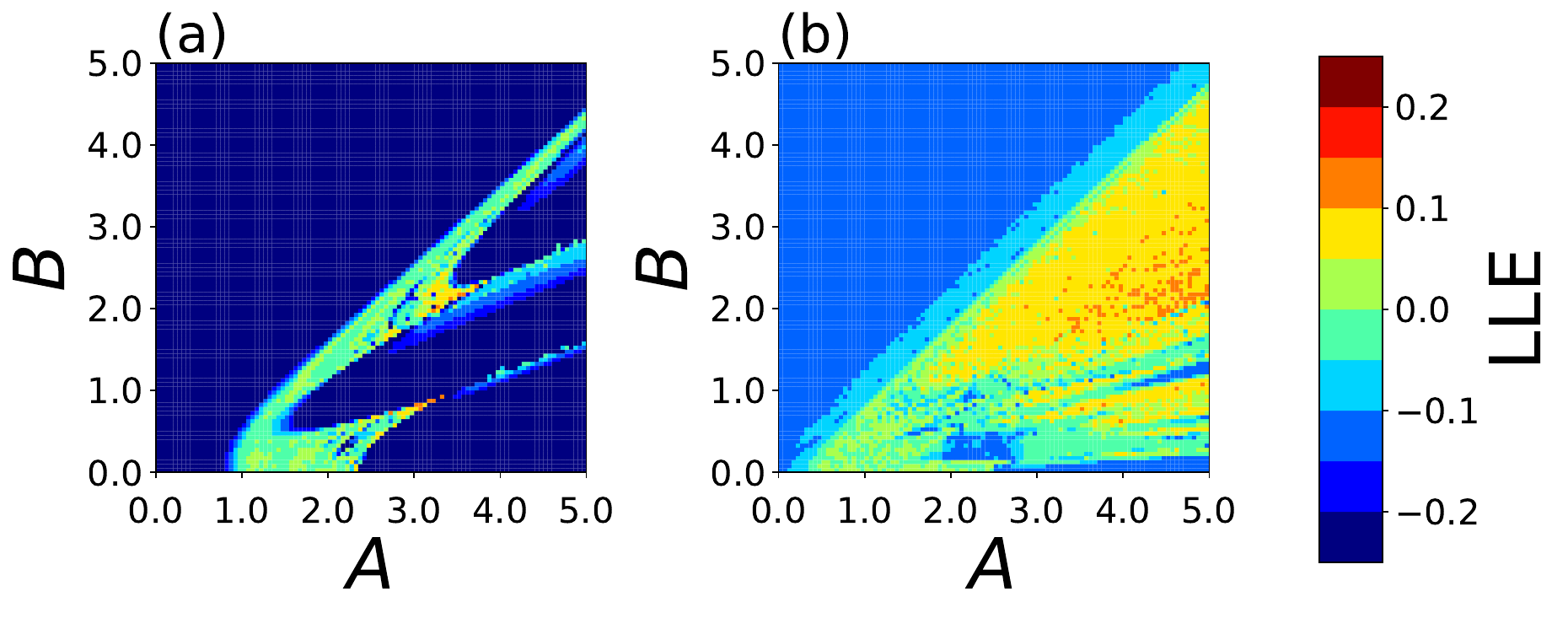}
\caption{Largest Lyapunov exponent  as a function of $A$ and $B$ in antisymmetric games ($C=-A, D=-B$). Colors from green to red indicate chaotic dynamics, while blue colors indicate convergence to a periodic attractor, which can be a limit cycle or a fixed point. In panel (b) players have longer memory.}
\label{fig:plot_lyapunov_exponents_AB}
\end{figure}

\noindent {\em Chaotic dynamics:}\\
 To check if the dynamics are chaotic or (quasi-)periodic, we consider a bifurcation diagram and calculate the Lyapunov exponents. In Fig. \ref{fig:plot_lyapunov_exponents_alpha} we fix one payoff matrix (we use the example of Fig. \ref{fig:TimeSeries_asymmetricgames_limitcycle}(d), i.e. $A=-C=-3.4$ and $B=-D=-2.5$) and set the sensitivity of choice to $\beta=1$. We then vary the memory-loss parameter $\alpha$. All fixed points are unstable for any $\alpha\in [0,1]$. In panel (a) we show the resulting bifurcation diagram. For each value of $\alpha$ we plot the coordinates $x$ the dynamics visits over the course of the trajectory, discarding an initial transient.  When there are only a few values of $x$, e.g. for $\alpha\in [0.4,0.5]$, the dynamics cycles between these values. When instead for a given value of $\alpha$ the dynamics visits significant portions of the phase space, as in $\alpha\in [0,0.2]$, the dynamics is chaotic. This is confirmed in panel (b) where we plot the largest Lyapunov Exponent  (LLE) $\lambda$; this exponent quantifies the exponential divergence of nearby trajectories  \citep{ott2002chaos}, positive values indicate chaotic dynamics.

Figure \ref{fig:plot_lyapunov_exponents_AB} shows that chaos is more frequently observed if the players have long memory. Indeed, in panel (b) we set $\alpha=0.01$, $\beta=1$, while in panel (a) it is $\alpha=0.7$. Chaos occupies a larger portion of the parameter space if one of the actions is dominant over the other, i.e. $B>0$, as opposed to the case $B=0$.
 The LLE is always negative if $B>A$, as the dynamics reaches a fixed point (consistently with the diagram depicted in Figure \ref{fig:ACnegative}). The LLE is larger for intermediate values of the payoffs, i.e. for large $A$ and $B$. 

\subsection{Arbitrary payoffs}
\label{sec:suppasymmgames}

\begin{figure}[h]
\centering
\includegraphics[width=0.8\textwidth]{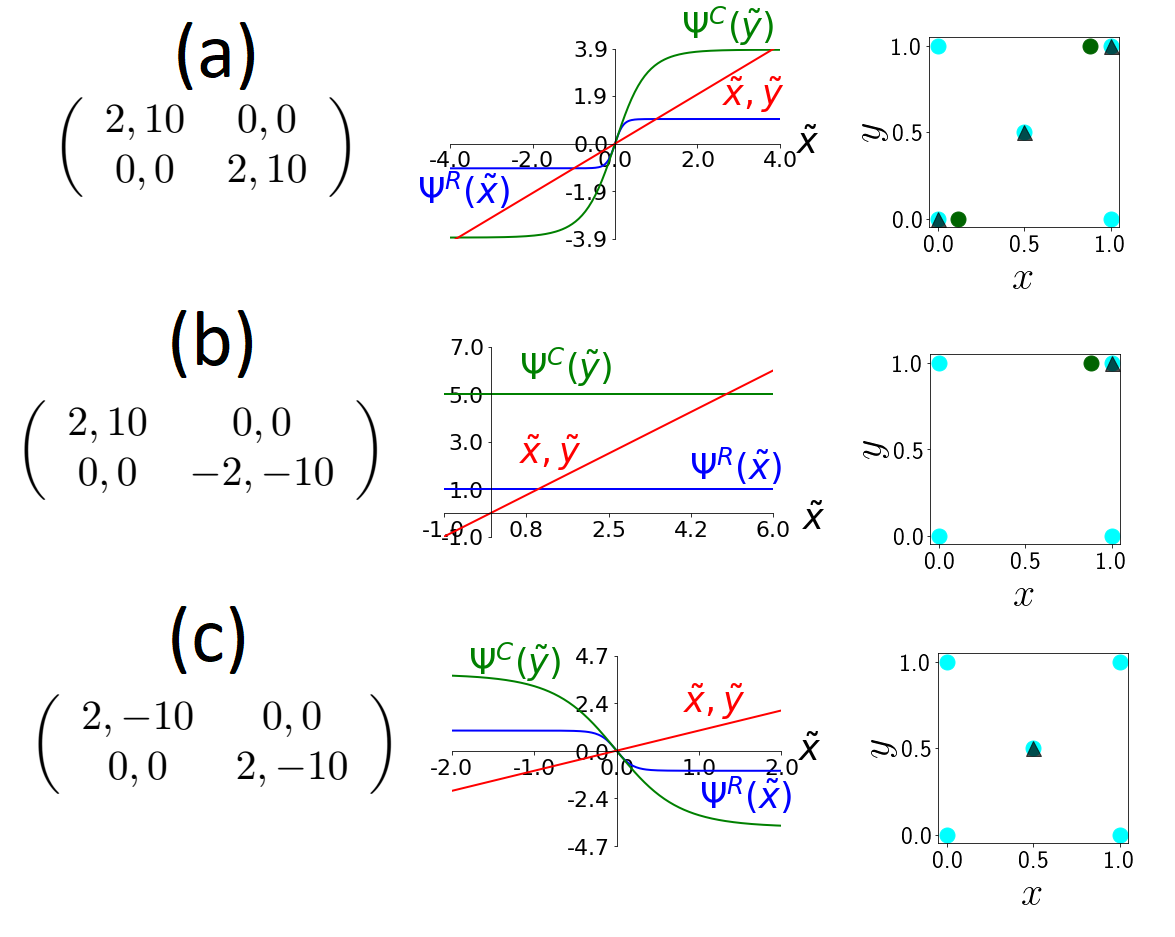}
\caption{Examples of asymmetric games in which $B\neq D$ and $A\neq C$. These games are analogous to games with constrained payoffs ($A=\pm C$ and $B=\pm D$) in the same class, i.e coordination, dominance-solvable and cyclic games respectively for panels (a) to (c). The only difference is that the player with highest payoffs --- and so strongest incentives --- plays a mixed strategy closer to the pure strategies.}
\label{fig:asymmetric}
\end{figure}

The most interesting difference between games with constrained payoffs and games with arbitrary payoffs occurs for games in which payoffs to one player are substantially larger than payoffs to the other player. Without loss of generality, consider the case in which payoffs to Column are much larger than payoffs to Row. In this case $D>>B$ and $C>>A$. Having larger payoffs, Column has strongest incentives to play better performing actions, and so he plays a mixed strategy closer to the pure strategies. We illustrate this with specific examples in Figure \ref{fig:asymmetric}, where we also show the functions $\Psi^R(\tilde{x}^\star)$ and $\Psi^C(\tilde{y}^\star)$. In panels (a) and (b) player Row has smaller payoffs, and so lower incentives. As a result, $x$ is always closer to the center of the strategy space than $y$. In panel (c) we show a similar payoff matrix to case (b) in Figure \ref{fig:ACnegative}, except that the large payoffs of player Column make the unique fixed point of \eqref{eq:fixedpoints-newcoordinates2} unstable.  

\subsection{Stochastic learning}
\label{sec:suppstochlearning}

\begin{figure}[h]
\centering
\includegraphics[width=0.85\textwidth]{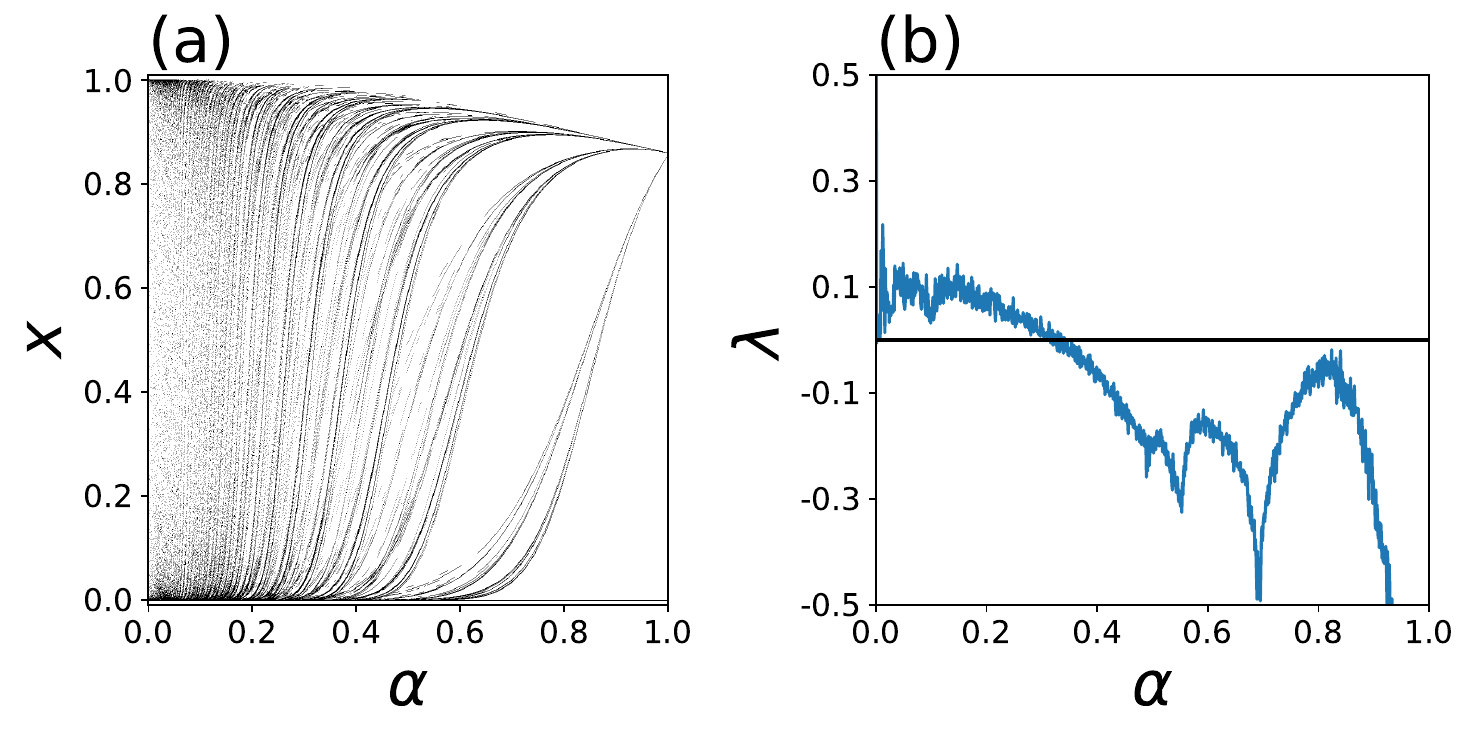}
\caption{Bifurcation diagram and largest Lyapunov exponent $\lambda$ as $\alpha$ is varied between 0 and 1. This figure is equivalent to Figure \ref{fig:plot_lyapunov_exponents_alpha}, except that here we consider stochastic learning. Chaos is robust to noise for small values of $\alpha$.}
\label{fig:plot_bifurcation_lyapunovexponent_stochastic}
\end{figure}

In Figure \ref{fig:plot_bifurcation_lyapunovexponent_stochastic} we show the bifurcation diagram and largest Lyapunov exponent as a function of $\alpha$ for stochastic learning. This figure is similar to Figure \ref{fig:plot_lyapunov_exponents_alpha}, consistently with theoretical studies on the effect of noise on dynamical systems \citep{crutchfield1982fluctuations}. The figure shows that chaos is robust to noise, as the LLE is positive for $\alpha\in [0,0.3]$. For $\alpha>0.6$ the dynamics only visits a few points, as can be seen in panel (a). This is because the players have short memories and so only a few different histories of actions played are possible. In the extreme case of no memory, $\alpha=1$, each player will ``jump'' between two points, corresponding to the two actions that her opponent may take at any time step. Indeed, in Figure \ref{fig:plot_lyapunov_exponents_alpha}(a) for $\alpha=1$ the dynamics only visits two points ($x=0$ and $x \approx 0.85$). This effect is absent in the deterministic dynamics, because the players choose distributions of actions.

\subsection{A few additional parameter combinations}
\label{sec:otherpars}

Here we cover some parameter and payoff combinations that have not been considered previously, and show that the results of our analysis can be directly applied to understand the learning behavior in these cases.

Consider the following dominance-solvable game
\begin{equation}
 \left( {\begin{array}{cc}
   1,6 & 5,-2 \\       3,2 & 1,-2     \end{array} } \right).
\label{eq:payoff2}
\end{equation}
Assume that $\delta=1$ (full consideration of forgone payoffs), that the dynamics is deterministic, and consider any value of $\alpha$, $\beta$ and $\kappa$. What dynamics can we expect? The payoff combinations are $A=-1.5$, $B=0.5$, $C=1$ and $D=3$. The payoffs do not satisfy any constraint of the type $A=\pm C$, $B=\pm D$. Differently from Section \ref{sec:suppasymmgames}, moreover, the payoffs to one player are not simply a rescaled version of the payoffs to the other player, so that the magnitude of the payoffs is not the only difference with respect to the baseline scenario. 

Nevertheless, we can qualitatively understand the outcome of learning based on our analysis. Because $A$ and $C$ have different signs, the functions $\Psi^R(\cdot)$ and $\Psi^C(\cdot)$ in Eq. \eqref{eq:fixedpoints-newcoordinates2} monotonically decrease, so there can only be one fixed point in the interior of the strategy space. The game is dominance solvable and, although both $B$ and $D$ have the same sign, the situation is similar to the upper left corner of the diagram in Figure \ref{fig:ACnegative}, where $B$ and $D$ had opposite sign. If $\alpha/\tilde{\beta}=\alpha/\{\beta \left[1-(1-\alpha)(1-\kappa) \right]\}$ is small, the fixed point is close to the unique pure NE of the game, located at $(s_2^R,s_1^C)$. If instead $\alpha/\tilde{\beta}$ is large, the fixed point is located in the center of the strategy space. Because $|D| > |B|$, finally, player Column always plays a strategy closer to the pure equilibrium than player Row, in line with the analysis in Section \ref{sec:suppasymmgames}.

Relax now the assumption $\delta=1$. Because we do not want to constrain $\alpha$ to take the value $\alpha=0$, the analysis in Section \ref{sec:reinforcement} does not directly apply. However, combining the effects of $\delta$ and $\alpha/\tilde{\beta}$ is straightforward. Positive values of $\alpha/\tilde{\beta}$ push all fixed points towards the center of the strategy space, so ``lock-in'' fixed points of the type discussed in Section \ref{sec:reinforcement} become less likely. 

\begin{figure}[h]
\centering
\includegraphics[width=1\textwidth]{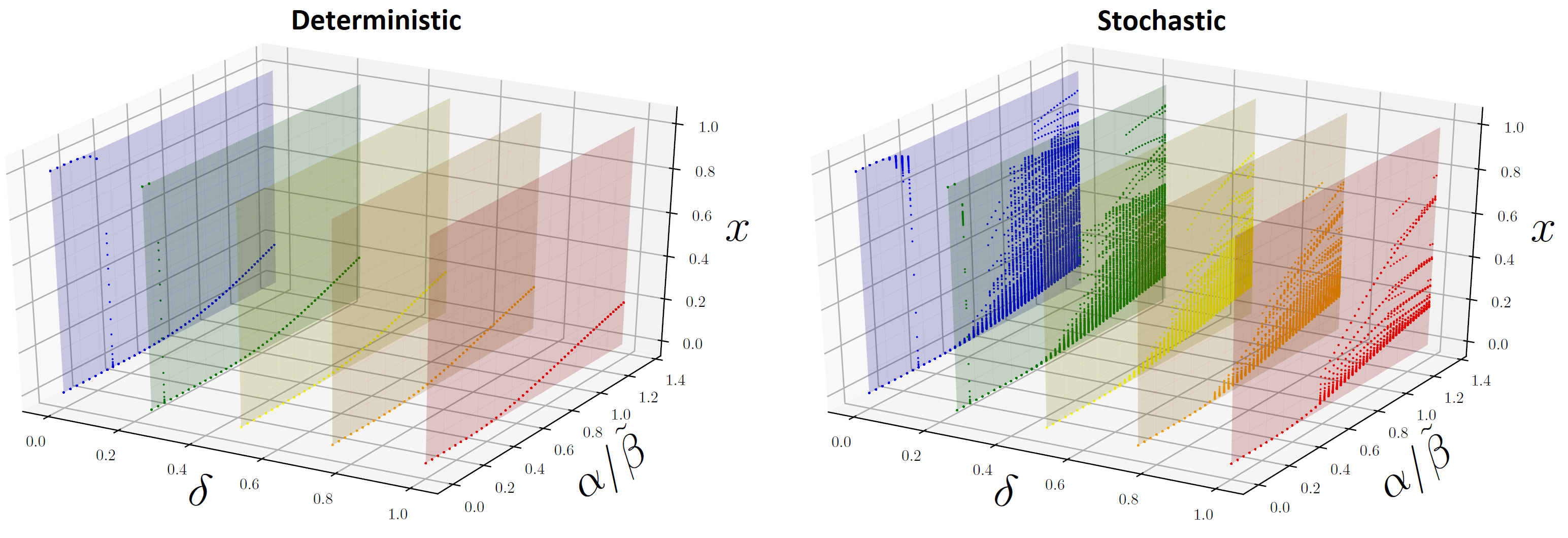}
\caption{Bifurcation diagram obtained by simulating the EWA equations for some values of the parameters $\delta$ and $\alpha/\tilde{\beta}$ (we only show $x$, the mixed strategy of player Row). Fixed points with $x=1$ are only possible for $\delta=0,0.2$ and for small values of $\alpha/\tilde{\beta}$.}
\label{fig:other_par_combinations}
\end{figure}

This is confirmed by simulating the EWA equations on the game in Eq. \eqref{eq:payoff2}, fixing $\kappa=0.5$, $\beta=0.5$, and varying $\delta$ and $\alpha$. We consider five values $\delta=0.00,0.25,0.50,0.75,1.00$, and several values of $\alpha\in[0,0.5]$. We simulate the dynamics and record the $x$ variables in the final time steps, obtaining the bifurcation diagram in Figure \ref{fig:other_par_combinations}. Consider the left panel, showing the deterministic dynamics. As $\alpha/\tilde{\beta}$ becomes larger, the fixed point moves towards the center of the strategy space, in line with the analysis in Section \ref{sec:baseline}. This is true for all values of $\delta$. However, for $\delta=0,0.2$, an additional fixed point at $x=1$ exists. This is one of the ``lock-in fixed points'' described in Section \ref{sec:reinforcement}, and it only exists for sufficiently small values of $\alpha/\tilde{\beta}$.

Finally, in Section \ref{sec:stoch} we have discussed robustness of our results to stochasticity, but we have fixed $\kappa$ and $\delta$ to $\kappa=1$ and $\delta=1$ respectively. The right panel of Figure \ref{fig:other_par_combinations} shows that robustness to stochasticity holds for the parameter values considered in this section, too. (The density of points in the bifurcation diagram indicates that the value of $x$ is often close to the deterministic one, although occasionally it is larger.)

While in this section we have not claimed that we can explicitly fully characterize the parameter space, we have shown a game and parameter combinations that had not yet been analyzed, but whose behavior could be qualitatively understood from the previous analysis.

\end{document}